\documentclass[amssymb,usenatbib,useAMS,usedcolumns]{mn2e}
\usepackage{times} 
\usepackage{latexsym,amssymb,amsmath,amsfonts}
\usepackage{rotfloat} 
\usepackage{rotating} 
\usepackage{float} 
\usepackage{graphicx}
\usepackage{natbib}
\usepackage{supertabular}
\usepackage{longtable}
\usepackage{url}
\usepackage{dcolumn}
\usepackage{placeins}
\usepackage{multirow} 
\usepackage{rotating}
\begin{document}
\newcommand{\sqcm}{cm$^{-2}$}  
\newcommand{\lya}{Ly$\alpha$}
\newcommand{\lyb}{Ly$\beta$}
\newcommand{\lyg}{Ly$\gamma$}
\newcommand{\heo}{\mbox{He\,{\sc i}}}
\newcommand{\he}{\mbox{He\,{\sc ii}}}
\newcommand{\hi}{\mbox{H\,{\sc i}}}
\newcommand{\hw}{\mbox{H\,{\sc ii}}}
\newcommand{\oth}{\mbox{O\,{\sc iii}}}
\newcommand{\ofo}{\mbox{O\,{\sc iv}}}
\newcommand{\of}{\mbox{O\,{\sc v}}}
\newcommand{\os}{\mbox{O\,{\sc vi}}}
\newcommand{\ose}{\mbox{O\,{\sc vii}}}
\newcommand{\oei}{\mbox{O\,{\sc viii}}}
\newcommand{\cf}{\mbox{C\,{\sc iv}}}
\newcommand{\cfi}{\mbox{C\,{\sc v}}}
\newcommand{\csi}{\mbox{C\,{\sc vi}}}
\newcommand{\cto}{\mbox{C\,{\sc ii}}}
\newcommand{\ct}{\mbox{C\,{\sc iii}}}
\newcommand{\sito}{\mbox{Si\,{\sc ii}}}
\newcommand{\sit}{\mbox{Si\,{\sc iii}}}
\newcommand{\sif}{\mbox{Si\,{\sc iv}}}
\newcommand{\sitw}{\mbox{Si\,{\sc xii}}}
\newcommand{\sfo}{\mbox{S\,{\sc iv}}}
\newcommand{\sfi}{\mbox{S\,{\sc v}}}
\newcommand{\ssi}{\mbox{S\,{\sc vi}}}
\newcommand{\nto}{\mbox{N\,{\sc ii}}}
\newcommand{\nt}{\mbox{N\,{\sc iii}}}
\newcommand{\nfo}{\mbox{N\,{\sc iv}}}
\newcommand{\nf}{\mbox{N\,{\sc v}}}
\newcommand{\pf}{\mbox{P\,{\sc v}}}
\newcommand{\nefo}{\mbox{Ne\,{\sc iv}}}
\newcommand{\nefi}{\mbox{Ne\,{\sc v}}}
\newcommand{\nesi}{\mbox{Ne\,{\sc vi}}}
\newcommand{\neo}{\mbox{Ne\,{\sc viii}}}
\newcommand{\nete}{\mbox{Ne\,{\sc x}}}
\newcommand{\mgt}{\mbox{Mg\,{\sc ii}}}
\newcommand{\fet}{\mbox{Fe\,{\sc ii}}}
\newcommand{\mgx}{\mbox{Mg\,{\sc x}}} 
\newcommand{\nani}{\mbox{Na\,{\sc ix}}} 
\newcommand{\arei}{\mbox{Ar\,{\sc viii}}} 
\newcommand{\alel}{\mbox{Al\,{\sc xi}}} 
\newcommand{\zabs}{$z_{\rm abs}$}
\newcommand{\zmin}{$z_{\rm min}$}
\newcommand{\zmax}{$z_{\rm max}$}
\newcommand{\zqso}{$z_{\rm qso}$}
\newcommand{\subHe}{_{\it HeII}}
\newcommand{\subH}{_{\it HI}}
\newcommand{\subHLy}{_{\it H Ly}}
\newcommand{\degree}{\ensuremath{^\circ}}
\newcommand{\lapp}{\mbox{\raisebox{-0.3em}{$\stackrel{\textstyle <}{\sim}$}}}
\newcommand{\gapp}{\mbox{\raisebox{-0.3em}{$\stackrel{\textstyle >}{\sim}$}}}
\newcommand{\be}{\begin{equation}}
\newcommand{\en}{\end{equation}}
\newcommand{\di}{\displaystyle}
\def\tworule{\noalign{\medskip\hrule\smallskip\hrule\medskip}} 
\def\onerule{\noalign{\medskip\hrule\medskip}} 
\def\bl{\par\vskip 12pt\noindent}
\def\bll{\par\vskip 24pt\noindent}
\def\blll{\par\vskip 36pt\noindent}
\def\rot{\mathop{\rm rot}\nolimits}
\def\alf{$\alpha$}
\def\refff{\leftskip20pt\parindent-20pt\parskip4pt}
\def\kms{km~s$^{-1}$}
\def\zabs{$z_{\rm abs}$}
\def\zem{$z_{\rm em}$}
\title[A new class of associated absorber]{
{\LARGE $HST/$COS} observations of a new population of associated QSO absorbers
\thanks{Based on observations made with the NASA/ESA {\sl Hubble Space Telescope}, obtained from the data archive at the Space Telescope Science Institute, which is operated by the Association of Universities for Research in Astronomy, Inc., under NASA contract NAS 5-26555.}}   
\author[S. Muzahid et al.]{S. Muzahid$^{1}$\thanks{E-mail: sowgat@iucaa.ernet.in},  
R. Srianand$^{1}$, N. Arav$^{2}$, B. D. Savage$^{3}$ and A. Narayanan$^{4}$ \\    
$^{1}$ Inter-University Centre for Astronomy and Astrophysics, Post Bag 4, 
Ganeshkhind, Pune 411\,007, India \\
$^{2}$ Department of Physics, Virginia Tech, Blacksburg, VA, 24061, USA \\   
$^{3}$ Department of Astronomy, University of Wisconsin, 475 North Charter Street, 
Madison, WI, 53706, USA \\ 
$^{4}$ Indian Institute of Space Science \& Technology, Thiruvananthapuram 695547, 
Kerala, India \\ 
}
\date{Accepted. Received; in original form}
\maketitle
\label{firstpage}
\begin {abstract}  
We present a sample of new population of associated absorbers, detected through \neo$\lambda\lambda$770,780 absorption, in $HST$/COS spectra of intermediate redshift (0.45 $< z < $ 1.21) quasars (QSOs). Our sample comprised of total 12 associated \neo\ systems detected towards 8 lines of sight (none of them are radio bright). The incidence rate of these absorbers is found to be 40\%. Majority of the \neo\ systems at small ejection velocities ($v_{\rm ej}$) show complete coverage of the background source, but systems with higher $v_{\rm ej}$ show lower covering fractions (i.e. $f_c \le 0.8$) and systematically higher values of $N(\neo)$. We detect \mgx\ $\lambda\lambda609,624$ absorption in 7 out of the 8 \neo\ systems where the expected wavelength range is cover by our spectra and is free of any strong blending. 
We report the detections of \nani $\lambda\lambda$681,694 absorption, for the first time, in three highest ejection velocity (e.g. $|v_{\rm ej}| \gtrsim 7,000$ \kms) systems in our sample. All these systems show very high $N(\neo)$ (i.e. $>10^{15.6}$~cm$^{-2}$), high ionization parameter (i.e. log~U~$\gtrsim$~0.5), high metallicity (i.e. $Z \gtrsim Z_{\odot}$), and ionization potential dependent $f_c$ values. The observed column density ratios  of different ions are reproduced by multiphase photoionization (PI) and/or collisional ionization (CI) equilibrium models. While solar abundance ratios are adequate in CIE, enhancement of $\rm Na$ relative to $\rm Mg$ is required in PI models to explain our observations.

The column density ratios of highly ionized species (i.e. \os, \neo, \mgx\ etc.) show a very narrow spread. Moreover, the measured $N(\neo)/N(\os)$ ratio in the associated absorbers is similar to what is seen in the intervening absorbers. All these suggest a narrow range of ionization parameter in the case of photoionization or a narrow temperature range (i.e. $T\sim 10^{5.9\pm0.1}$ K) in the case of CIE models. The present data does not distinguish between these two alternatives. However, detection of absorption line variability with repeat $HST/$COS observations will allow us to (i) distinguish between these alternatives, (ii) establish the location of the absorbing gas and (iii) understand the mechanism that provides stability to the multiphase medium. These are important for understanding the contribution of associated \neo\ absorbers to the AGN feedback.   
\end {abstract} 
\section{Introduction} 
\label{sec_intro}  

Associated absorbers are unique tools to probe the physical conditions of the gaseous environment in the immediate vicinity of the background quasar (QSOs). The abundance of heavy elements in these absorbers provides a direct measure of the star formation and chemical evolution in the center of galaxies hosting QSOs \citep[]{Hamann97a}. Most importantly, good fraction of associated absorbers are believed to originate from the ejected material from the central engine of the QSOs \citep[]{Gordon99}. These outflows are theoretically invoked to regulate the star formation of the host galaxies and growth of the supper massive black holes (SMBHs) at their centers \citep[]{Silk98,King03,Bower06,Ostriker10}.   

There is no firm definition in the literature for an associated absorber. The absorbers with velocity spread of few $\times$ 100 \kms, which appear within few $\times$ 1000 \kms\ from the emission redshift of the QSO, are generally defined to be associated absorbers \citep{Hamann97a}. In addition to the proximity to the QSOs, associated absorbers are also characterized by (a) time variable absorption line strength \citep[]{Barlow89,Barlow92,Hamann95,Srianand01,Hall11,Vivek12} (b) very high metallicity \citep[e.g. near solar abundances;][]{Petitjean94a,Hamann97a} and high ionization parameter \citep[e.g. log~U $\gtrsim 0.0$;][]{Hamann98,Hamann00,Muzahid12b} (c) partial coverage of the continuum source  \citep[e.g.][]{Barlow97,Srianand99,Ganguly99,Arav08} and (d) presence of excited fine structure lines \citep[e.g.][]{Srianand00apm}. These above mentioned properties are unlikely to occur in intervening systems \citep[however see][for a very special case]{Balashev11} and thus, they are believed to originate from gas very close to the QSO or a possible ejecta of the central engine. In the case of gas outflowing from the QSO, line driven radiative acceleration has often been suggested to be an important driving mechanisms however, only a handful of convincing evidences exists in the literature till date \citep[see e.g.,][]{Arav94,Arav95,Srianand02}. 

Based on their line widths, the associated absorbers are broadly classified into two categories: (1) the broad absorption line (BAL) and (2) narrow absorption line (NAL) systems.
BALs and NALs are predominantly detected through species (e.g., \mgt, \cf, \sif, \nf\ etc.) with low ionization potential (i.e. IP $\lesssim 100$ eV) in the UV-optical regime. On the other hand, the soft X-ray spectra of $\sim$ 40 -- 50\% of the Seyfert galaxies and QSOs happen to show K-shell absorption edges of highly ionized oxygen \citep[i.e., \ose, \oei\ with IP $\gtrsim 0.5$~keV;][]{Reynolds97a,George98,Crenshaw03}, known as X-ray ``warm absorbers" (WAs). These X-ray WAs are often said to correlate with the presence of absorption in the UV regime \citep[see e.g.,][]{Mathur94,Mathur95b,Mathur95a,Mathur98,Mathur99,Brandt00,Arav07}. \citet{Telfer98} have argued that the BAL-like absorption seen towards SBS~1542$+$541 could be a potential X-ray WA candidate \citep[see also][]{Hamann95}. However, QSOs known to have associated BAL absorption are generally found to be X-ray weak \citep[]{Green95,Green96,Stalin11}. In few cases the physical conditions in the UV absorbers are shown to be incompatible with that of X-ray WAs \citep[e.g.,][]{Srianand00apj,Hamann00}.  Therefore, although a unified picture of X-ray and UV associated absorbers is desirable, it is not clear whether there is any obvious connection between them. Even in cases, where simultaneous occurrences of the X-ray and UV absorption are seen, the absorbing gas need not be co-spatial. For example, envisaging a disk-wind model, \cite{Murray95b} have shown that the X-ray absorption originates very close to the accretion disk whereas UV absorption predominantly occurs in the accelerated gas farther away.   

The study of the species with ionization potential intermediate between UV-optical and X-ray absorbers (i.e., few $\times$ 100 eV) is important to understand the comprehensive nature of the ionization structure and thus the unified picture of QSO outflows detected in different wavebands. The resonant transitions of highly ionized species (e.g. \neo, \nani, \mgx, \alel\ and \sitw), that fall in the far-ultraviolet (FUV)  regime, are ideally suited for studying the intermediate ionization conditions of the associated absorbers. However, only a handful of absorbers showing some of these species have been reported till date. For example, the first tentative detection of associated \neo\ absorption was reported by \citet{Korista92} towards Q~0226$-$1024. The three other tentative detections existing in literature are by \citet{Petitjean96} towards HS~1700$+$6414, \citet{Gupta05} towards 3C~48 and \citet{Ganguly06} towards HE~0226$-$4110. We also note that a possible \neo\ detection is reported in the composite {\sl Far-Ultraviolet Spectroscopic Explorer} ($FUSE$) spectrum by \citet{Scott04}. There are only six confirmed detections of \neo\ absorption in associated absorbers reported till date [i.e., UM~675, \citet{Hamann95}; SBS~1542$+$541, \citet{Telfer98}; PG~0946$+$301, \citet{Arav99a}; J2233$-$606, \citet{Petitjean99}; 3C~288.1, \citet{Hamann00}; HE~0238$-$1904, \citet{Muzahid12b}]. Among these, both SBS~1542$+$541 and PG~0946$+$301 are BALQSOs. While multiphase photoionization models are generally used to explain most of these observations, \citet{Muzahid12b} have shown that the models of collisional ionization equilibrium can also reproduce the observed column density ratios of high ions like \os, \neo\ and \mgx. Therefore the collisional ionization could be an equally important ionizing mechanism in these absorbers.  

Although the highly ionized UV absorbers are of prime interest to probe the ``missing link" between UV and X-ray continuum absorbers, the low  rest frame wavelengths (i.e. $\lambda_{\rm rest}<$ 912 \AA) of the diagnostic species (e.g. \neo, \nani, \mgx\ etc.) make them difficult to detect. This is partly because of the Galactic Lyman-limit absorption in the spectra of low redshift sources and the \lya\ forest contamination in the spectra of high redshift sources. Hence the intermediate redshift (i.e. 0.5 $<$ \zem\ $<$ 1.5) UV bright QSOs are ideal for this study. Note that such a study is only feasible with far-ultra-violet (FUV) sensitive space based telescopes like {\sl Hubble Space Telescope} ($HST$). In this paper we present a sample of new class of associated absorbers detected through \neo\ absorption in the FUV spectra of intermediate redshift QSOs obtained with the {\sl Cosmic Origins Spectrograph} (COS) on board $HST$.   

This paper is organized as follows. In Section~2 we describe the observations and data reduction techniques for the sample of QSOs studied here. In Section~3 we discuss the effects of partial coverage in column density measurements and how we correct for it. Data sample and analysis of individual absorption systems are presented in Section~4. In Section~5 we explore ionization models for some of these systems detected with number of different ions. In Section~6 we discuss the overall properties of these absorbers. We summarize our main results in Section~7.  
Throughout this paper we use flat $\Lambda$CDM cosmology with ($\Omega_{\rm M}$, $\Omega_{\Lambda}$) = (0.27,0.73) and a Hubble parameter of $H_{0}$ = 71 \kms Mpc$^{-1}$. The solar relative abundances of heavy elements are taken from \citet{Asplund09}.

%

%
\begin{table*} 
\vfill  
\begin{center}  
\caption{Details of observations of lines of sight where we search for associated \neo\ absorption.} 
\begin{tabular}{lllccccr}  
\hline 
No. & QSO   & \zem  & Grating  & $t_{\rm exp}$ (ks)   & Coverage (\AA)  & Prop. ID & PI     \\  
(1)   &  (2)  &  (3)     &     (4)          &    (5)          &   (6)    & (7)  & (8)  \\ 
\hline 
\hline 

1. & 3C 263$^{\ast}$         &   $0.646$ \citep[]{Marziani96}   &   G130M   &   $15.3$   &   $1140 - 1450$  &   11541   &   Green \\
   &                         &                                  &   G160M   &   $18.0$   &   $1405 - 1790$  &           &         \\
2. & FBQS 0751+2919          &   $0.916$  (SDSS)                &   G130M   &   $16.5$   &   $1160 - 1465$  &   11741   &   Tripp \\
   &                         &                                  &   G160M   &   $23.2$   &   $1410 - 1795$  &           &         \\
3. & HB89 0107--025          &   $0.956$  \citep[]{Surdej86}    &   G130M   &   $21.2$   &   $1160 - 1465$  &   11585   & Crighton\\
   &                         &                                  &   G160M   &   $21.1$   &   $1410 - 1795$  &           &         \\
4. & HB89 0232--042$^{\ast}$ &   $1.440$  \citep[]{Janknecht06} &   G130M   &   $16.0$   &   $1160 - 1465$  &   11741   &   Tripp \\
   &                         &                                  &   G160M   &   $22.8$   &   $1410 - 1795$  &           &         \\
5. & HE 0153--4520           &   $0.451$  \citep[]{Wisotzki00}  &   G130M   &   $5.2$    &   $1140 - 1450$  &   11541   &   Green \\
   &                         &                                  &   G160M   &   $5.9$    &   $1405 - 1790$  &           &         \\
6. & HE 0226--4110           &   $0.493$  \citep[]{Ganguly06}   &   G130M   &   $6.7$    &   $1140 - 1450$  &   11541   &   Green \\
   &                         &                                  &   G160M   &   $7.7$    &   $1405 - 1790$  &           &         \\
7. & HE 0238--1904           &   $0.629$  \citep[]{Muzahid12b}  &   G130M   &   $6.4$    &   $1140 - 1450$  &   11541   &   Green \\
   &                         &                                  &   G160M   &   $7.4$    &   $1405 - 1790$  &           &         \\
8. & HS 1102+3441            &   $0.509$  (SDSS)                &   G130M   &   $11.3$   &   $1140 - 1450$  &   11541   &   Green \\
   &                         &                                  &   G160M   &   $11.2$   &   $1405 - 1790$  &           &         \\
9. &LBQS 0107--0235$^{\ast}$ &   $0.957$   \citep[]{Surdej86}   &   G130M   &   $28.2$   &   $1140 - 1455$  &   11585   & Crighton\\  
   &                         &                                  &   G160M   &   $44.4$   &   $1410 - 1795$  &           &         \\ 
10.&LBQS 1435--0134$^{\ast}$ &   $1.310$  (SDSS)                &   G130M   &   $22.3$   &   $1160 - 1465$  &   11741   &   Tripp \\
   &                         &                                  &   G160M   &   $34.1$   &   $1410 - 1795$  &           &         \\
11.&PG 1148+549              &   $0.975$  (SDSS)                &   G130M   &   $17.8$   &   $1160 - 1465$  &   11741   &   Tripp \\
   &                         &                                  &   G160M   &   $18.4$   &   $1410 - 1795$  &           &         \\
12.&PG 1206+459              &   $1.163$  (SDSS)                &   G130M   &   $17.3$   &   $1160 - 1465$  &   11741   &   Tripp \\
   &                         &                                  &   G160M   &   $36.1$   &   $1410 - 1795$  &           &         \\
13.&PG 1259+593              &   $0.476$  (SDSS)                &   G130M   &   $9.2$    &   $1140 - 1450$  &   11541   &   Green \\
   &                         &                                  &   G160M   &   $11.1$   &   $1405 - 1790$  &           &         \\
14.&PG 1338+416              &   $1.214$  (SDSS)                &   G130M   &   $22.7$   &   $1160 - 1465$  &   11741   &   Tripp \\
   &                         &                                  &   G160M   &   $35.0$   &   $1410 - 1795$  &           &         \\
15.&PG 1407+265              &   $0.940$ \citep[]{McDowell95}   &   G130M   &   $16.6$   &   $1160 - 1465$  &   11741   &   Tripp \\
   &                         &                                  &   G160M   &   $17.3$   &   $1410 - 1795$  &           &         \\
16.&PG 1522+101              &   $1.328$  (SDSS)                &   G130M   &   $16.4$   &   $1160 - 1465$  &   11741   &   Tripp \\
   &                         &                                  &   G160M   &   $23.0$   &   $1410 - 1795$  &           &         \\
17.&PG 1630+377              &   $1.475$  (SDSS)                &   G130M   &   $22.9$   &   $1160 - 1465$  &   11741   &   Tripp \\
   &                         &                                  &   G160M   &   $14.3$   &   $1410 - 1795$  &           &         \\
18.&PKS 0405--123$^{\ast}$   &   $0.573$ \citep[]{Laor94}       &   G130M   &   $22.1$   &   $1140 - 1450$  &   11508, 11541 &   Noll, Green \\
   &                         &                                  &   G160M   &   $11.0$   &   $1405 - 1790$  &           &         \\
19.&PKS 0552--640$^{\ast}$   &   $0.680$ \citep[]{Grazian02}    &   G130M   &   $9.2$    &   $1140 - 1425$  &   11692   &   Howk  \\
   &                         &                                  &   G160M   &   $8.2$    &   $1400 - 1745$  &           &         \\
20.&PKS 0637--752$^{\ast}$   &   $0.653$ \citep[]{Hunstead78}   &   G130M   &   $9.6$    &   $1140 - 1425$  &   11692   &   Howk  \\
   &                         &                                  &   G160M   &   $8.6$    &   $1400 - 1745$  &           &         \\
\hline 
\hline 
\end{tabular} 
\end{center}  
Notes -- Column 2 and 3 list the name and emission redshift of the QSOs respectively. In the parenthesis we provide the references for emission redshift measurements. Column 4 lists the FUV gratings used for the observations. Column 5 is the total exposure time in kilo-seconds for each grating setting. Column 6 is the total wavelength coverage for the choice of grating. Column 7 lists the $HST$ ID of the proposal for which the observations were carried out and; Column 8 lists the PI of the proposal. All data were retrieved from the {\sl Mikulski Archive for Space Telescopes} ($MAST$) and reduced using the {\sc CalCOS} pipeline v.2.12. $^{\ast}$Sources with 5 GHz flux density excess of 50 mJy.    
\label{tab:data} 
\end{table*}

\section{Observations and data reduction}  
\label{sec_obs}

The sample in which we searched for \neo\ absorbers had the following selection criteria: (1) archived $HST$/COS FUV spectra (G130M+G160M) of quasars which were public as of February 2012, (2) QSOs with emission redshift $z_{\rm em} \ge 0.45$ so that the the \neo\ doublet transitions (770\AA\ and 780\AA) are redshifted into the wavelength coverage of the COS G130M and G160M gratings, and (3) spectra with signal-to-noise ratio ($S/N)$ per resolution element $>$10.     
The properties of COS and its in-flight operations are discussed by \citet{Osterman11} and \citet{Green12}. The data were retrieved from the $HST$ archive and reduced using the STScI {\sc CalCOS} v.2.12 pipeline software. The reduced data were flux calibrated. The alignment and addition of the separate G130M and G160M exposures were done using the software developed by the COS team\footnote{http://casa.colorado.edu/$\sim$danforth/science/cos/costools.html}. The exposures were weighted by the integration time while coadding in flux units. The procedures followed for data reduction are described in greater detail in \citet{Narayanan11}. 
The unabsorbed QSO continuum is fitted using low-order polynomials interpolated between wavelength ranges devoid of strong absorption lines. We use the standard procedure that propagates the continuum placement uncertainty to the normalized flux. 

The medium resolution ($R \sim$ 20,000) with $S/N \ge 10$ COS data, covering 1150 -- 1800 \AA\ wavelength range, allow us to search for \neo $\lambda\lambda$770,780 doublets in the redshift range $\sim$~0.45$-$1.31. Observational details of our final sample of 20 quasar sight lines are listed in Table~\ref{tab:data}. Half of these sight lines were part of a blind survey to detect the warm-hot intergalactic gas (prop. ID 11741). Of the remaining, majority are from the COS-GTO program (prop. ID 11541) to probe the gas phases in the low redshift IGM and galaxy halos. For the QSO PKS~0405$-$123, we have combined spectra obtained under the GTO program of the COS science team from December 2009 and the $HST$ Early Release Observations (prop. ID 11508) of August 2009. While weak radio emission (i.e. flux density $\le$ 1 mJy) is detected in most of the QSOs in our sample, only seven of them (called radio bright from now on) have radio flux density in excess of 50 mJy at 5 GHz.

%
\begin{figure} 
\centerline{
\vbox{
\centerline{\hbox{ 
\includegraphics[height=4.4cm,width=8.8cm,angle=00]{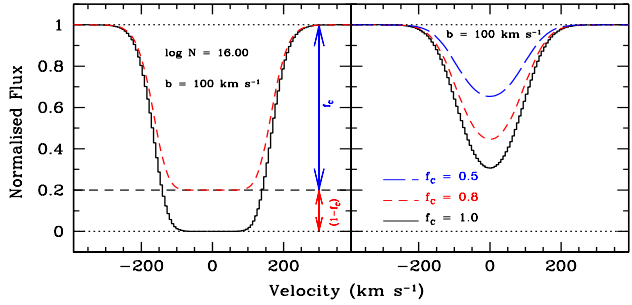} 
}}
}}
\caption{Demonstration of the effect of partial coverage in case of a heavily saturated 
({\sl left}) and an unsaturated ({\sl right}) \neo\ lines. Partially covered saturated 
line will appear as flat bottom profile with nonzero flux. For a partially covered 
unsaturated line, column density measurement can lead to much lower value compared to the 
true value if we do not correct for the covering fraction.         
} 
\label{fc_fun} 
\end{figure} 
%

\section{Partial Coverage and Uncertainty in Column density measurement}  
\label{sec_parcov}

Because of the close physical association between the background QSO and the associated absorber, in many cases it so happens that the latter does not cover the former entirely. In such cases, the observed residual intensity at any frequency can be written as, 
\be 
I(\nu) = I_{0}(\nu)(1-f_{c})+f_{c} I_{0}(\nu) {\rm exp}[-\tau(\nu)]~ . 
\label{eqn:covf1}
\en 
Here $I_{0}(\nu)$ is the incident intensity, $\tau(\nu)$ is the true optical depth, and $f_{c}$ is the covering fraction. In the case of doublets with rest frame wavelengths $\lambda_{1}$ \& $\lambda_{2}$ and oscillator strengths $f_{1}$ \& $f_{2}$, the residual intensities $R_{1}$ and $R_{2}$ in the normalized spectra, at any velocity $v$ with respect to the line centroid are related by  
\be 
R_{2}(v) = (1-f_{c})+f_{c}\times \left(\frac{R_{1}(v)-1+f_{c}}{f_{c}}\right)^{\gamma},  
\label{eqn:covf2}
\en  
where $\gamma = f_{2}\lambda_{2}/f_{1}\lambda_{1}$. The value of $\gamma$ is very close to 2 for doublets  \citep[see e.g.,][]{Srianand99,Petitjean99}. This equation in principle allows us to calculate the covering fraction of the absorbing gas.    
%

\begin{table}
\caption{List of important Extreme-UV (EUV) lines used in this paper$^{1}$.} 
\centering  
\begin{tabular}{crrrcc} 
\hline 
 Ion & IP(1)$^{a}$ & IP(2)$^{b}$ & $\lambda^{c}$ (\AA) & $f_{\rm osc}^{d}$ & log~$T_{\rm max}^{e}$  \\   
 (1) &       (2)   &     (3)     &    (4)              &    (5)       & (6)       \\  
\hline 
\ofo   &    54.93    &    77.41 &   787.7105 &  1.11$\times10^{-1}$ & 5.20 \\ 
       &             &          &   608.3968 &  6.70$\times10^{-2}$ &      \\ 
\of\   &    77.41    &   113.90 &   629.7320 &  5.15$\times10^{-1}$ & 5.40 \\   
\nfo\  &    47.45    &    77.47 &   765.1467 &  6.16$\times10^{-1}$ & 5.15 \\  
\nefi\ &    97.12    &   126.22 &   572.3380 &  7.74$\times10^{-2}$ & 5.45 \\  
\nesi\ &   126.22    &   157.93 &   558.5940 &  9.07$\times10^{-2}$ & 5.65 \\ 
\neo\  &   207.28    &   239.10 &   770.4089 &  1.03$\times10^{-1}$ & 5.85 \\  
       &             &          &   780.3240 &  5.05$\times10^{-2}$ &      \\ 
\arei\ &   124.32    &   143.45 &   700.2450 &  3.85$\times10^{-1}$ & 5.75 \\ 
       &             &          &   713.8100 &  1.88$\times10^{-1}$ &      \\        
\nani\ &   264.19    &   299.88 &   681.7190 &  9.24$\times10^{-2}$ & 5.90 \\ 
       &             &          &   694.1460 &  4.54$\times10^{-2}$ &      \\   
\mgx\  &   328.24    &   367.54 &   609.7930 &  8.42$\times10^{-2}$ & 6.05 \\ 
       &             &          &   624.9410 &  4.10$\times10^{-2}$ &      \\   
\alel\ &   399.37    &   442.08 &   550.0310 &  7.73$\times10^{-2}$ & 6.15 \\ 
       &             &          &   568.1200 &  3.75$\times10^{-2}$ &      \\  
\sitw$^{\dagger}$ & 476.08  &  523.52 &   499.4060 &  7.19$\times10^{-2}$ & 6.35 \\ 
                  &         &         &   520.6650 &  3.45$\times10^{-2}$ &      \\  
\hline 
\end{tabular}
~\\ 
\flushleft  
$^{1}$From \citet{Verner94} \\ 
$^{a}$Creation ionization potential \\  
$^{b}$Destruction ionization potential \\ 
$^{c}$Rest frame wavelength in \AA \\ 
$^{d}$Oscillator strength  \\ 
$^{e}$Temperature at which collisional ionization fraction \citep[]{Sutherland93} peaks \\ 
$^{\dagger}$Not covered for any of the systems reported here  
\label{tab:atomic_data} 
\end{table}    

The effects of partial coverage in case of a heavily saturated and an unsaturated line are shown in Fig.~\ref{fc_fun}. In the left panel of the figure we plot synthetic profiles of \neo$\lambda$770 line (true line center optical depth $\tau_0$= 21.0) with $N(\neo)$ = 10$^{16}$ cm$^{-2}$ and $b$-parameter of 100 \kms\ for $f_c = 1.0$ (solid profile) and $f_c = 0.8$ (dashed profile). The heavy saturation in the profile with complete coverage suggests large optical depth (i.e. $e^{-\tau(\nu)}$ = 0). The dashed curve showing flat bottom profile but flux level not reaching to zero, clearly suggests a partial coverage scenario with $f_c = 1 - I(\nu)/I_{0}$. Evidently, presence of only one line is sufficient to compute $f_c$ in such a situation.     
In the right hand panel of Fig.~\ref{fc_fun}, we show synthetic profiles of \neo$\lambda$770 line ($\tau_0$= 2.1) with $N$ = 10$^{15}$ cm$^{-2}$ and $b$-parameter of 100 \kms\ for $f_c = 1.0$ (solid), $f_c = 0.8$ (short dashed) and  $f_c = 0.5$ (long dashed). For the same column density, profiles with different covering fraction look different. The line center becomes shallower for lower $f_c$ values. The line with a column density of 10$^{15}$ cm$^{-2}$ will appear as $N(\neo)$ = 10$^{14.85}$ cm$^{-2}$ ($\tau_0$= 1.5) and 10$^{14.60}$ cm$^{-2}$ ($\tau_0$= 0.8) for covering fractions of $f_c =$ 0.8 and 0.5 respectively. Evidently, the observed optical depth in this case is degenerate between the true optical depth and the covering fraction. Therefore, unlike the saturated case, we need at least two lines from the same ground state to estimate the true column density. After estimating the covering fraction (either by flat bottom approximation or from doublet transitions) for a given species we recover the true optical depth by inverting Eq.~\ref{eqn:covf1}. We then use the partial coverage corrected flux for Voigt profile fitting. Here we make an explicit assumption that the individual Voigt profile components in a blend all have same $f_c$ for a given ion, 
but note that $f_c$ can be strongly dependent on the velocity along the absorption trough \citep[]{Srianand99,Arav99b,Gabel05,Arav08}. In addition inhomogeneous absorption models were shown to produce good fits for the absorption troughs as well \citep[]{Arav08,Borguet12a}. However, given the survey nature of this work and the limited $S/N$ of the data, we deem it adequate to treat the absorber with simple covering fraction models and to reserve the use of more elaborate models for future investigations and high $S/N$ observations.   

\section{Data Sample and Analysis}  
\label{sec_data} 
\begin{table*}
\begin{sideways} 
\begin{minipage}{22cm} 
\begin{center}  
\caption{Summary of properties of the associated \neo\ absorbers.} 
\begin{tabular}{lccrclcccccc} 
\hline 
 QSO                      &   \zem  &  \zabs    & $v_{\rm ej}$ & log~$L_{912{\rm \AA}}$ & log~$N(\hi)$  & log~$N(\os)$ & log~$N(\neo)$ & log~$N(\nani)$ & log~$N(\mgx)$     & Comments & $\delta v$  \\  
 &         &           &   (\kms)      &  (in erg s$^{-1}$ Hz$^{-1}$) & ($N$ in cm$^{-2}$) &  ($N$ in cm$^{-2}$) &  ($N$ in cm$^{-2}$) &  ($N$ in cm$^{-2}$) & ($N$ in cm$^{-2}$) &  & (\kms) \\ \hline 
  (1)                     &    (2)  &      (3)  &  (4)          &  (5)  &    (6)          &    (7)    &   (8)      &   (9)         &    (10)  &    (11)  &   (12)       \\ 
\hline \hline 
HE~0226$-$4110            &  0.493  &  0.49272  &   $-56$       & 31.24 & 14.49$\pm$0.01    &  14.76$\pm$0.13  &  14.09$\pm$0.19  &  NA		&  NA	           & Secure    & 226.1   \\ 
HS~1102+3441              &  0.509  &  0.48518  &   $-4768$     & 30.39 & $<$14.52          &  15.00$\pm$0.18  &  15.22$\pm$0.20  &  NA	        &  NA   	   & Secure    & 661.3   \\ 
HE~0238$-$1904            &  0.629  &  0.59795  &   $-5767$     & 31.46 & $<$13.71          &  $<$13.6         &  14.22$\pm$0.03  &  $<$14.19	        &  $<$14.82         & Tentative & 337.3   \\  
                          &         &  0.60406  &   $-4624$     &       & $<$14.45          &  15.24$\pm$0.09  &  15.62$\pm$0.02  &  $<$14.79	        &  $>$ 15.32\footnotemark[1]       & Secure    & 435.5   \\  
                          &         &  0.60989  &   $-3538$     &       & $<$14.60          &  15.06$\pm$0.07  &  15.50$\pm$0.22  &  $<$13.90	        &  $>$ 15.11\footnotemark[1]	   & Secure    & 460.3   \\  
FBQS~0751+2919            &  0.916  &  0.91983  &   $+598$      & 31.74 & $<$14.34          &  NA	       &  14.59$\pm$0.09  &  $<$13.94        &  BL	         & Secure    & 171.0   \\  
PG~1407+265               &  0.940  &  0.93287  &   $-1103$     & 31.90 & $<$13.71          &  NA	       &  14.36$\pm$0.22  &  $<$13.60        &  14.29$\pm$0.18  & Secure    & 96.0    \\  	
HB89~0107$-$025           &  0.956  &  0.94262  &   $-2057$     & 31.16 & $<$14.75          &  NA	       &  14.27$\pm$0.04  &  $<$13.75        &  BL 	           & Tentative & 117.6   \\  
PG~1206+459               &  1.163  &  1.02854  &   $-19228$    & 32.09 & $\sim$14.00\footnotemark[2]  &  NA		&  $>$16.11	     &  15.12$\pm$0.09        &  15.59$\pm$0.08  & Secure    & 358.9   \\  
PG~1338+416               &  1.214  &  1.15456  &   $-8156$     & 31.48 & $<$13.88          &  NA           &  $>$16.05        &  15.25$\pm$0.05	&  15.80$\pm$0.04  & Secure    & 337.9   \\  
                          &         &  1.16420  &   $-6818$     &       & BL                &  NA	    &  $\sim$15.62     &  $>$15.47$\pm$0.17  &  $\sim$15.47       & Secure    & 474.9   \\  
                          &         &  1.21534  &   $+181$      &       & 14.04$\pm$0.04\footnotemark[3]    &  14.74$\pm$0.12\footnotemark[3] &  14.42$\pm$0.05  &  $<$14.18    &  14.62$\pm$0.06 & Secure           & 225.6   \\ 
\hline \hline 
\label{tab_list} 
\end{tabular}
\end{center}  
Notes -- Column~1 lists the QSO sight lines in which signatures of associated \neo\ absorption is detected. Note that the presence of associated \neo\ absorbers towards HE~0226--4110 and HE~0238--1904 have been reported previously by \citet{Ganguly06} and \citet{Muzahid12b} respectively. Column~2 lists the emission redshifts of the QSOs; Column~3 lists the \neo\ optical depth weighted redshifts of the absorbers. Column~4 lists the ejection/outflow velocities of the \neo\ absorbers, defined as the velocity separation between the \zem\ and \zabs. Column~5 lists the luminosities of the QSOs at the rest frame 912\AA. Column~6,7,8,9,10 list the measurements/limits on integrated column densities of \hi, \os, \neo, \nani\ and \mgx\ respectively. During the column density estimations, effects of the partial coverage have been taken care of, whenever applicable. The species that are not available in the COS spectrum are marked by `NA'. The species that are heavily blended are marked by `BL'.  Column~11 tells us whether the \neo\ detection is secure. Column~12 lists the line spreads of \neo\ absorption. To compute the line spread we followed the procedure as described in \cite[][see their Fig.~3]{Muzahid12a}.                
~\\ 
\footnotemark[1] {Detected in $FUSE$} ~\\  
\footnotemark[2] {Detected in $HST/$STIS (E230M)} ~\\  
\footnotemark[3] {Detected in $HST/$FOS (G270H)} ~\\ 
\end{minipage}  
\end{sideways} 
\end{table*} 
%
 

%
\begin{figure} 
\centerline{
\vbox{
\centerline{\hbox{ 
\includegraphics[height=8.4cm,width=9.2cm,angle=00]{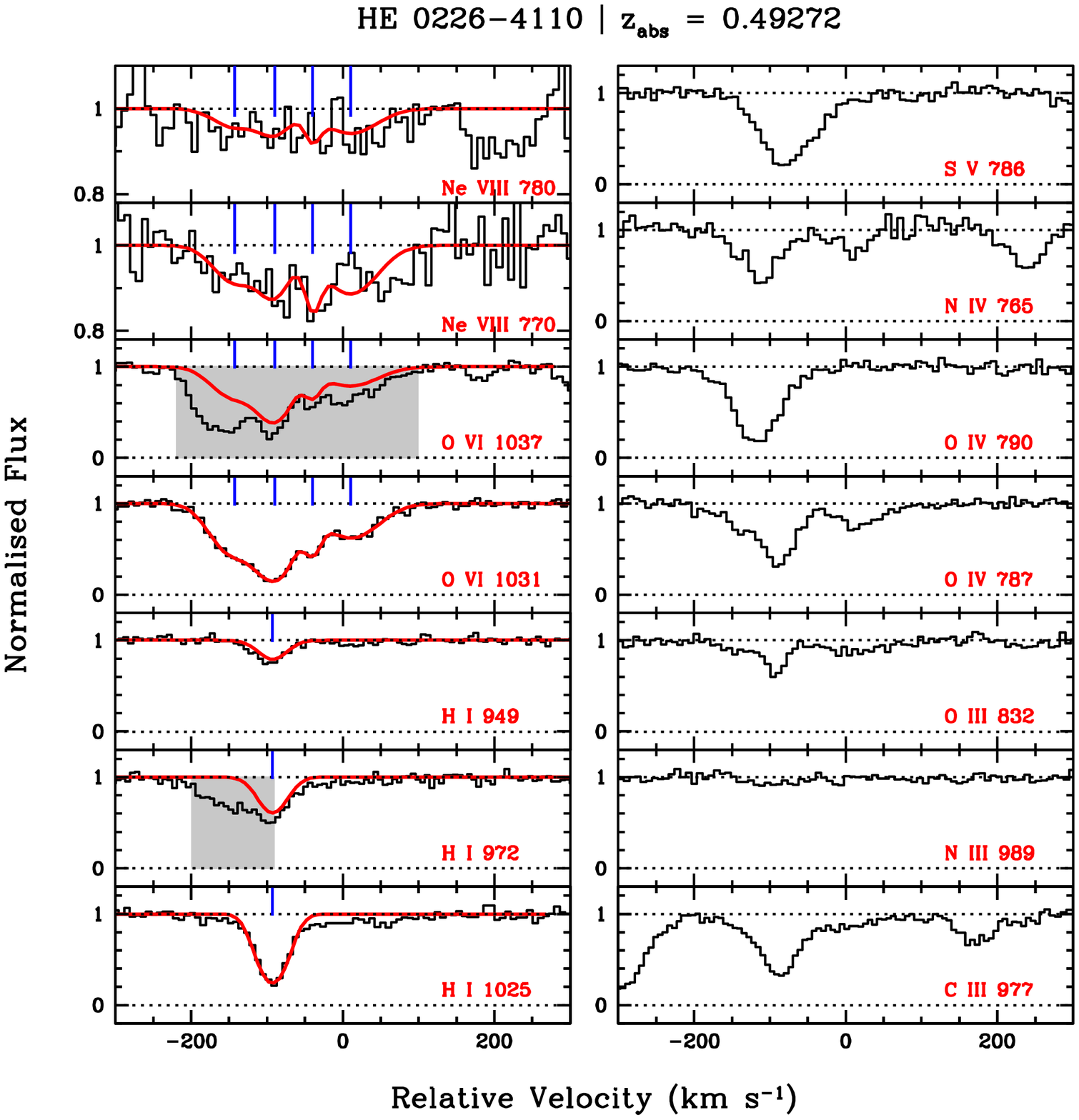} 
}}
}}
\caption{Velocity plot of the associated \neo\ absorption system at \zabs\ = 0.49272 
towards HE~0226$-$4110. The zero velocity corresponds to the emission redshift 
(\zem\ = 0.493) of the QSO. In case of \hi, \os\ and \neo, the smooth curves are the 
best fitting Voigt profiles, overplotted on top of the data. The vertical ticks mark 
the centroids of the individual Voigt profile components. Absorption features unrelated 
to this system are marked by the shaded regions.} 
\label{vp_he0226} 
\end{figure} 
%

For the analysis presented in this paper, we concentrate on associated absorbers detected through \neo$\lambda\lambda$770,780 doublets. In Table~\ref{tab:atomic_data} we have summarised some of the important EUV lines used in this paper including \neo. Here we define, an associated absorbers as (a) those with ejection velocities $|v_{\rm ej}|$ $\lesssim$ 8000 \kms\ with respect to the QSO emission redshift \citep[see e.g.][]{Fox08}, or (b) show clear signatures of partial coverage (see e.g. Section \ref{sec_parcov}) even when having higher ejection velocities (i.e. $|v_{\rm ej}|$~$\ge 8000$ \kms). Here, the ejection velocity $v_{\rm ej}$ is defined as the velocity separation between the emission redshift of the QSO and the \neo\ optical depth weighted redshift of the absorber. The $-ve$ sign in the ejection velocity is used whenever absorber redshift is less than the emission redshift of the QSO (i.e. \zabs\ $\le$ \zem). However, in subsequent discussions we will use the term ``higher velocity" assuming modulus of the ejection velocity.   

We have searched for the \neo\ doublets in the relevant spectral range by imposing the doublet matching criteria. For each identified coincidences we checked the consistency of the profile shape. However, we do not impose the condition of optical depth ratio consistency for the \neo\ doublets, keeping in mind the effects of partial coverage as discussed in Section \ref{sec_parcov}. We then checked for the presence of all other species at the redshift of the identified \neo\ doublets.  We find the signatures of associated \neo\ absorption only in 8 out of 20 (40\%) lines of sight. We have detected 12 associated \neo\ absorption systems in total towards 8 lines of sight. Note that any continuous absorption comprised of single/multiple component(s) are treated as system. Apart from the system detected towards PG~1206+459, all other systems are detected within $\sim$8000 \kms\ with respect to the QSOs. Because of clear signature of partial coverage in the \neo\ doublet we have included the system in our sample. Based on the number per unit redshift of \neo\ absorbers \citep{Narayanan09} we expect to detect only 2 \neo\ absorbers from the intervening gas. Interestingly none of these \neo\ absorption detected is towards the 7 radio bright QSOs. Although we search up to 8000 \kms, 67 per cent of the absorbers are detected within 5000 \kms\ from the emission redshift of the QSO.   

Details of the sight lines and the \neo\ absorbers are summarized in Table~\ref{tab_list}. Apart for \zabs\ = 0.94262 towards HB89~0107$-$025 (marked as ``Tentative" in column \#11 of Table~\ref{tab_list}), all other associated system in our sample show at least one other species which indeed makes our \neo\ identification robust. Next, we provide details of each individual \neo\ systems detected in our sample.  

\subsection{\zabs = 0.49272 towards HE~0226$-$4110}  
\label{sec_discript_HE0226_0.49272}

The ejection velocity of the absorber is only $\sim -$56 \kms. The velocity plot of this system clearly shows that the \neo\ absorption is spread over $\sim$~226 \kms\ (see Fig.~\ref{vp_he0226}). However, as \neo\ doublets occur in the extreme blue end of the COS spectrum, the $S/N$ is not high. A tentative detection of \neo\ in this system in the $FUSE$ data was reported earlier by \citet{Ganguly06}. Here we confirm their detection. Apart from the weak \neo, other ions detected in the COS spectrum are \ct, \oth, \nfo, \ofo, \os\ and possibly \sfi. The detection of \of\ is also reported in the $FUSE$ spectrum by \citet{Ganguly06} which is not covered by the COS spectrum.  
As the \os~$\lambda 1037$ line is severely blended (see Fig.~\ref{vp_he0226}), we could not use \os\ doublets to estimate the \os\ covering fraction. \neo, on the other hand, is very weak. \nani\ and \mgx\ doublets as well as \lya\ line are not covered by the COS spectrum. Nevertheless, unblended profiles of Ly$\beta$ and Ly$\delta$ transitions are found to be consistent with covering fraction ($f_c$) being 1.0, suggesting complete occultation of the background source by the absorber. The measured column density is log~$N(\hi)$ [cm$^{-2}$] = 14.49$\pm$0.01. The unblended \os\ $\lambda 1031$ profile is fitted with four Voigt profile components. The total column density (i.e. the summed up column densities measured in four components) is log~$N(\os)$ [cm$^{-2}$] = 14.76$\pm$0.13.  Because of the low $S/N$ ratio we use the component structure of \os\ absorption to fit the \neo\ doublets keeping the $b$-parameter tied with the corresponding \os\ component. The estimated total column densities of \neo\ is log~$N(\neo)$ [cm$^{-2}$] = 14.09$\pm$0.19. The total column densities for \neo\ and \os\ as reported by \citet{Ganguly06}, using apparent optical depth technique, (i.e. log~$N(\neo)$ [cm$^{-2}$] = 14.25$\pm$0.15 and log~$N(\os)$ [cm$^{-2}$] = 14.84$\pm$0.08) are very similar to our measurements. The difference in profile between high ions (e.g. \os, \neo) and low ions (e.g. \hi, \ct, \oth, etc.) is clearly evident from the system plot. Only the strongest \os\ component is accompanied  by these low ions. Such a difference in profiles possibly suggests multiphase nature of the absorbing gas. A detailed discussion on the absorbing system and the QSO properties can be found in \citet{Ganguly06}, therefore we do not discuss this system in detail.     

%
\begin{figure} 
\centerline{
\vbox{
\centerline{\hbox{ 
\includegraphics[height=8.4cm,width=9.0cm,angle=00]{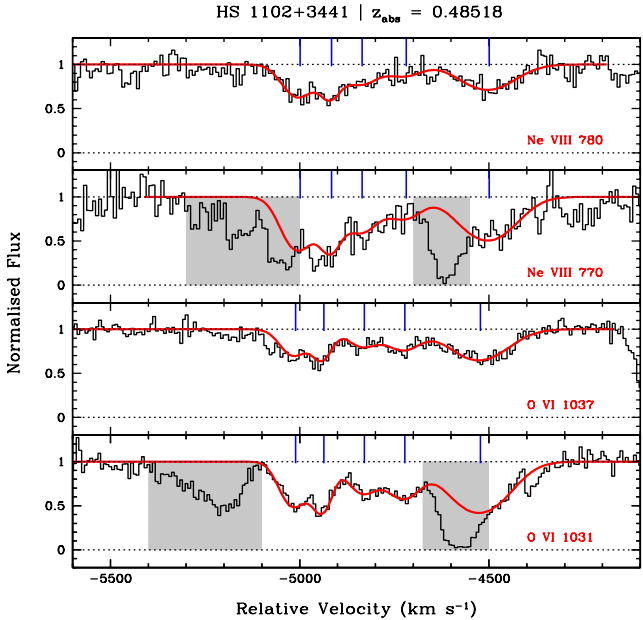} 
}}
}}
\caption{Velocity plot of the associated \neo\ absorption system at \zabs\ = 0.48518 
towards HS~1102$+$3411. The zero velocity corresponds to the emission redshift 
(\zem\ = 0.509) of the QSO. The smooth curves overplotted on top of the data are the 
best fitting Voigt profiles. The vertical ticks mark the centroids of the individual 
Voigt profile components. Absorption features unrelated to this absorber are marked by 
the shaded regions. 
}  
\label{vp_hs1102} 
\end{figure} 
%

\subsection{\zabs = 0.48518 towards HS~1102$+$3441} 
\label{sec_discript_HS1102_0.48518}

The ejection velocity of this system is $v_{\rm ej} \sim -4768$ \kms\ and is 
detected through \os\ and \neo\ absorption, kinematically spread over $\sim$700 \kms\ (see Fig.~\ref{vp_hs1102}). The \neo$\lambda$770 is blended with unknown contaminants whereas \os\ $\lambda 1031$ is found to be blended with \lya\ absorption of \zabs\ = 0.26165 system. Unblended profiles of \neo\ $\lambda 780$ and \os\  $\lambda 1037$ clearly show multicomponent structures with at least five components contributing to the absorption. Because of the blending we do not attempt to estimate the covering fraction for either of the detected ions. The Voigt profile fitting assuming complete coverage seems to give reasonably good fit to the unaffected pixels of the blended profiles. The estimated total column densities are log~$N(\os) [{\rm cm^{-2}}] = 15.00 \pm 0.18$ and log~$N(\neo)[{\rm cm^{-2}}] = 15.22 \pm 0.20$. \lya\ is not covered by the COS spectrum. \lyb\ and \lyg\ lines are  contaminated. Nevertheless, we use the contaminated \lyb\ profile to put an upper limit on $N(\hi)$. Assuming component structure and  $b$-parameters similar to \neo, we find $N(\hi)< 10^{14.52}$~cm$^{-2}$. 

%

\subsection{\zabs = 0.59795, 0.60406 \& 0.60989 towards HE~0238$-$1904}  
\label{sec_discript_HE0238}  

These systems are detected at $v_{\rm ej} \sim -4500$ \kms\ away from the emission redshift of the QSO, in seven absorption components kinematically spread over $\sim$~1800 \kms. We have presented a detailed analysis of this absorber in an earlier paper \citep[see][]{Muzahid12b}. \mgx\ lines from this system are severely affected by the Galactic H$_{2}$ absorption and we were able to measure $N(\mgx)$ only in some of the components showing \neo\ detection. The \nani\ doublets are not covered by the COS spectrum for this system. We looked at $FUSE$ LiF2a data covering the \nani\ doublets but do not find any clear signature of \nani\ absorption.


\begin{figure} 
\centerline{
\vbox{
\centerline{\hbox{ 
\includegraphics[height=8.4cm,width=8.4cm,angle=00]{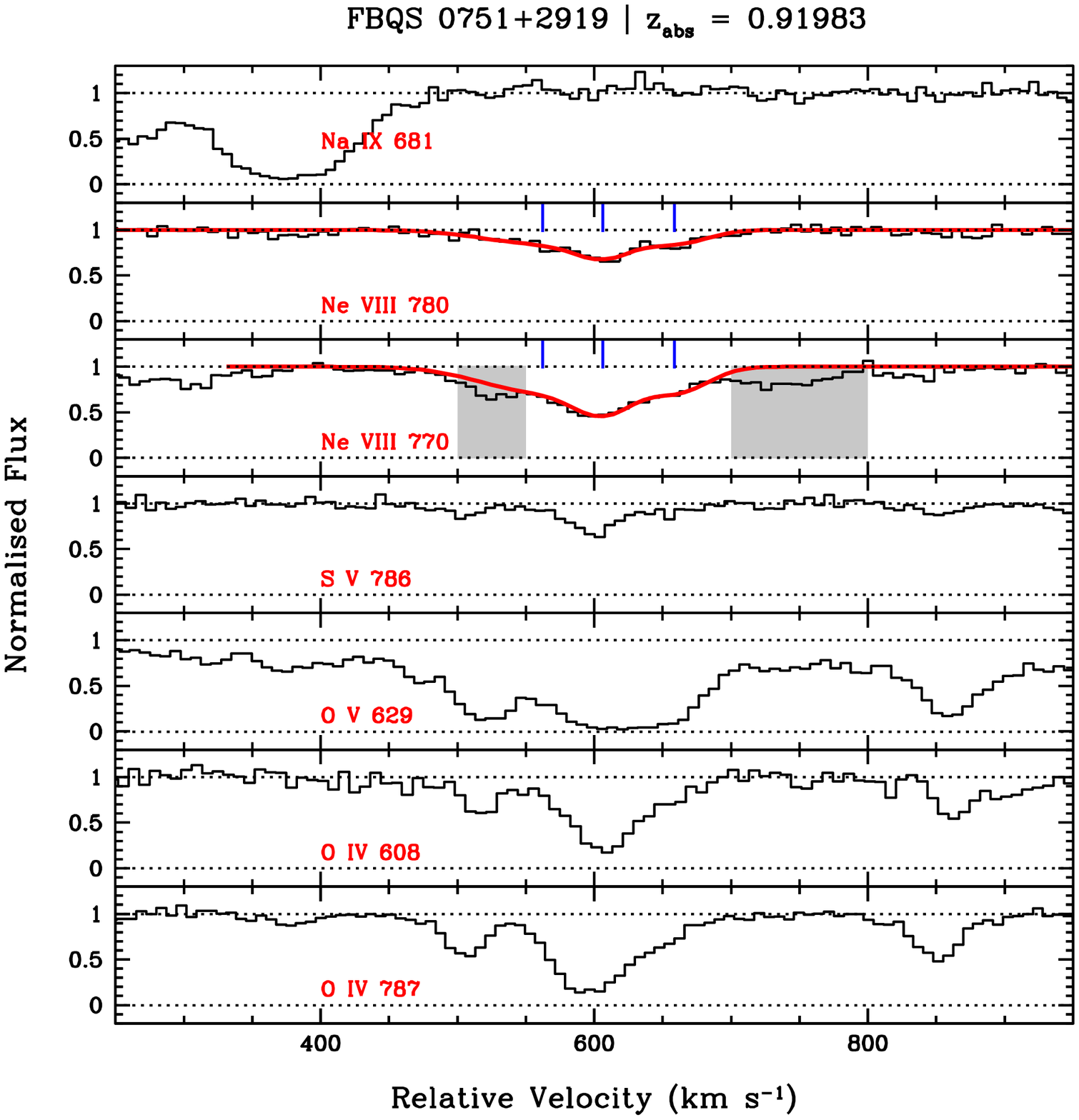} 
}}
}}
\caption{Velocity plot of the associated \neo\ absorption system at \zabs\ = 0.91983 
towards FBQS~0751$+$2919. The zero velocity corresponds to the emission redshift 
(\zem\ = 0.915) of the QSO. The smooth curves overplotted on top of the data in the 
\neo\ panel are the best fitting Voigt profiles. The vertical ticks mark the centroids 
of the individual Voigt profile components.} 
\label{vp_fqbs0751} 
\end{figure} 

\subsection{\zabs = 0.91983 towards FBQS~0751$+$2919} 
\label{sec_discript_FBQS0751_0.91983}

The ejection velocity of this system is $v_{\rm ej} \sim +598$ \kms, suggesting \zabs~$>$~\zem. \neo\ doublets in this system clearly show multicomponent structure spreads over $\sim$~170~\kms\ (see Fig.~\ref{vp_fqbs0751}). Apart from \neo, other ions detected in this system are \ofo, \of\ and \sfi. \neo$\lambda$770 seems to be mildly blended in both the wings. The optical depth ratios in the core pixels of \neo\ absorption are consistent with $f_c$=1.0. The Voigt profile fitting of the \neo\ doublets leads to a total column density of log~$N(\neo) [{\rm cm^{-2}}] = 14.59 \pm 0.09$.  For this system \os\ lines are not covered by the COS spectrum. The clear non-detection of \nani\ $\lambda681$ transition is consistent with log~$N(\nani)[{\rm cm^{-2}}]<$ 13.94 at 3$\sigma$ confidence level. The expected positions of both the members of \mgx\ doublet are heavily blended and hence we do not have any estimate on $N(\mgx)$. Very high order Lyman series lines (i.e. with $\lambda_{\rm rest}<$930 \AA) are covered by the COS spectrum where we do not find any clear signature of \hi\ absorption. Non-detection of Ly$-9$ transition is consistent with $N(\hi)<10^{14.34}$~cm$^{-2}$. 

\subsection{\zabs = 0.93287 towards PG~1407$+$265}   
\label{sec_discript_PG1407_0.93287}  
%

\begin{figure} 
\centerline{
\vbox{
\centerline{\hbox{ 
\includegraphics[height=8.4cm,width=8.4cm,angle=00]{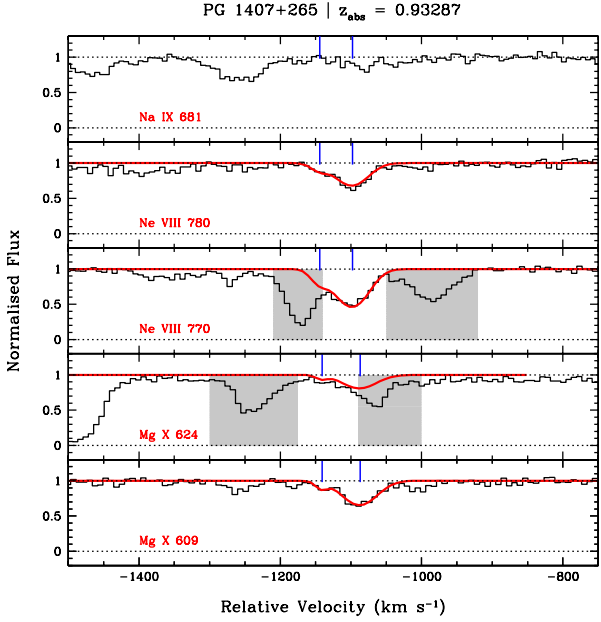} 
}}
}}
\caption{Velocity plot of the associated \neo\ absorption system at \zabs\ = 0.93287 
towards PG~1407$+$265. The zero velocity corresponds to the emission redshift (\zem\ = 0.940) 
of the QSO. The smooth curves overplotted on top of the data are the best fitting Voigt 
profiles. The vertical tick marks the centroids of the individual Voigt profile components.} 
\label{vplot_PG1407} 
\end{figure} 
%
The ejection velocity of this system is $v_{\rm ej} \sim -1103$ \kms. \neo\ absorption has two components spread over $\sim$~100~\kms\ (see Fig.~\ref{vplot_PG1407}). \neo$\lambda$770 line is found to be contaminated in both the wings. Nevertheless, the unblended core pixels are consistent with covering fraction $f_c = 1.0$. We estimate log~$N(\neo)[{\rm cm^{-2}}] = 14.36\pm0.22$. \mgx\ doublet is fitted with two components slightly off-centered with respect to the \neo\ components. Estimated total column density of \mgx\ absorption is log~$N(\mgx)[{\rm cm^{-2}}] = 14.29\pm0.18$. The non-detection of \nani\ $\lambda 681$ transition is consistent with log~$N(\nani)[{\rm cm^{-2}}]<13.60$ at 3$\sigma$ confidence level. \os\ doublets are not covered by the COS spectrum. 
Very high order Lyman series lines (i.e. with $\lambda_{\rm rest}<$ 930 \AA) are covered by the COS spectrum where we do not find any clear signature of \hi\ absorption. In addition, no convincing \lyb\ (or \lyg) absorption is seen in archival $HST$/FOS spectrum, obtained with the G190H grating. We note that the non-detection of \lyb\ is consistent with $N(\hi)<10^{13.71}$~cm$^{-2}$.    

\subsection{\zabs = 0.94262 towards HB89~0107--025}  
\label{sec_discript_HB890107_0.94262}  

\begin{figure} 
\centerline{
\vbox{
\centerline{\hbox{ 
\includegraphics[height=8.4cm,width=8.4cm,angle=00]{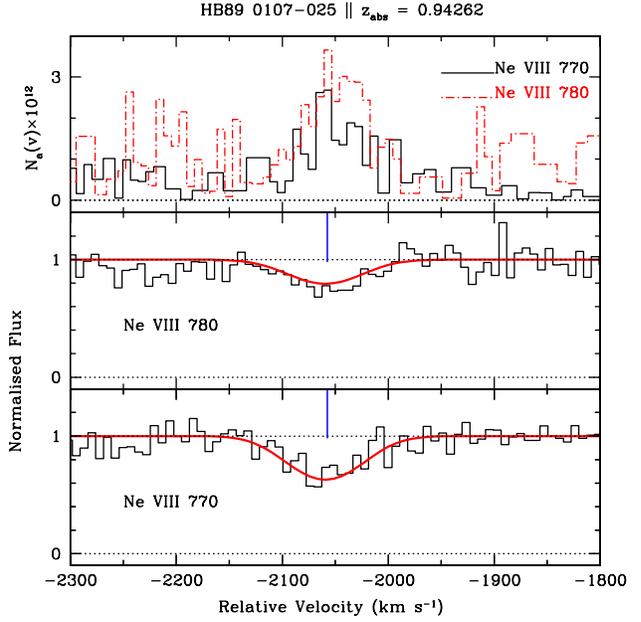} 
}}
}}
\caption{Velocity plot of the associated \neo\ absorber at \zabs\ = 0.94262 
towards HB89~0107--025. The zero velocity corresponds to the emission redshift 
(\zem\ = 0.956) of the QSO. The smooth curves overplotted on top of the data are 
the best fitting Voigt profiles. The vertical tick marks the line centroid. The 
apparent column density profiles of \neo\ doublets 
[in units of 10$^{12}$ cm$^{-2}$(\kms)$^{-1}$] are plotted in the top panel.}  
\label{aod_hb890107} 
\end{figure}

The ejection velocity of this system is $v_{\rm ej} \sim -2057$~\kms\ and is detected only through \neo\ absorption spread over $\sim 120$~\kms\ (see Fig.~\ref{aod_hb890107}). The covering fraction of \neo\ is consistent with $f_c = 1$ within the continuum placement uncertainty. We measure log~$N(\neo)$[cm$^{-2}$] = $14.27\pm0.04$, whereas the non-detection of \nani~$\lambda 681$ transition in the COS spectrum is consistent with $N(\nani) < 10^{13.75}$~cm$^{-2}$ at 3$\sigma$ confidence level. The expected positions of \mgx\ doublets are contaminated and thus we cannot confirm its presence. In addition, we do not detect any other ion in the COS spectrum, corresponding to this system. \os\ is not covered by COS and we do not find any signature of \os\ lines  in the FOS/G190H spectra. Therefore, we treat this system as tentative one. The non-detection of \lyb\ in FOS/190H spectra is consistent with $N(\hi)< 10^{14.75}$~cm$^{-2}$.

\subsection{\zabs = 1.02854 towards PG~1206$+$459} 
\label{sec_discript_PG1206_1.02854}

%
\begin{figure} 
\centerline{
\vbox{
\centerline{\hbox{ 
\includegraphics[height=10.4cm,width=9.4cm,angle=00]{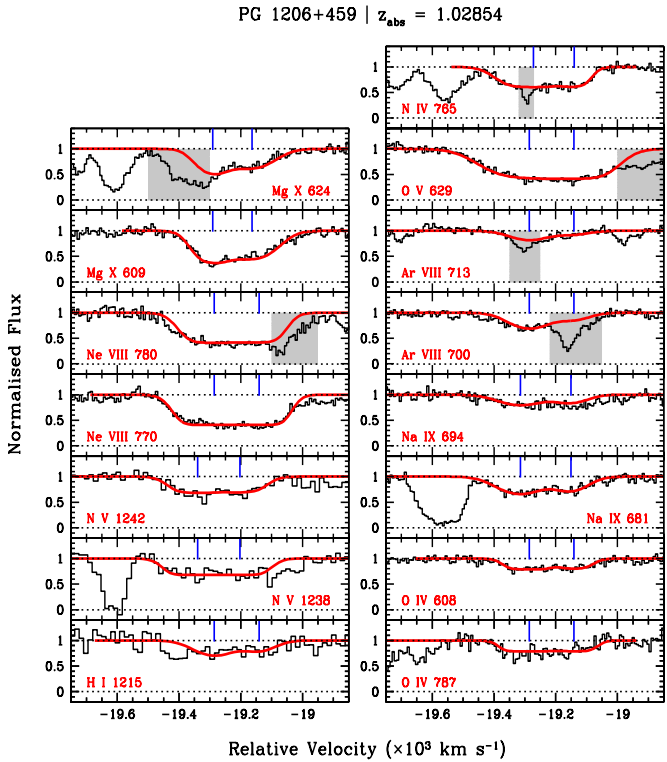} 
}}
}}
\caption{Velocity plot of the associated \neo\ absorption system at \zabs\ = 1.02854 
towards PG~1206$+$459. The zero velocity corresponds to the emission redshift 
(\zem\ = 1.163) of the QSO. The smooth curves overplotted on top of the data are the 
best fitting Voigt profiles after correcting for the effect of partial coverage. 
The vertical ticks mark the centroids of the individual Voigt profile components.   
\lya\ and \nf\ are from STIS E230M spectrum. Absorption lines unrelated to 
this system are marked by the shaded regions.} 
\label{vp_pg1206} 
\end{figure} 

This is the highest ejection velocity associated system detected in our sample, with $v_{\rm ej}$ $\sim -19,228$ \kms, and \neo\ absorption is spread over $\sim360$~\kms. This system is part of our sample, despite having large ejection velocity, as it shows clear signature of partial coverage. In Fig.~\ref{vp_pg1206} we show absorption profiles of different species as a function of their outflow velocity with respect to the QSO emission redshift (\zem\ = 1.214). The highly ionized species like \arei$\lambda\lambda$700,713; \neo$\lambda\lambda$770,780; \nani$\lambda\lambda$681,694 and \mgx$\lambda\lambda$609,624, originating from this absorber are detected in the COS spectrum. In addition, we also detect species like \ofo, \of, \nfo\ in COS and \hi, and \nf\ in the $HST$/STIS E230M spectrum. The STIS spectrum does not cover \lyb, \ct\ or \cf\ lines. However, expected wavelength range of \sif, \sit~$\lambda$1206, \sito~$\lambda$1260 and \cto~$\lambda$1334 lines are covered in the STIS data, but we do not detect any of these species. The profiles of \ofo, \nfo, \nf\ and \neo\ doublets are flat over $\sim$~300 \kms, indicating partial coverage and heavy saturation of these lines. The flat bottom assumption (see Section~\ref{sec_parcov}) gives the covering fractions for \ofo, \nfo, \nf, and \neo\ as $f_c$ = 0.21, 0.40, 0.32 and 0.59 respectively. Unlike these species, the doublets of \mgx\ absorption are unsaturated. The uncontaminated profile of \mgx\ $\lambda 609$ clearly shows component structure with at least two components contributing to the absorption. For the subsequent discussions on this system (see section~\ref{sec_phot_model_pg1206}) we will refer the higher and lower velocity components (i.e. blue and red) as component-1 and component-2 respectively. The blue wing of the \mgx\ $\lambda 624$ is blended with S~{\sc iv} $\lambda 657$ line from \zabs\ = 0.9275. The core pixels of \mgx\ $\lambda 624$ which are not affected by this blending are consistent with $f_c = 0.68$. For the singlet transition of \of, we have taken $f_c$ = 0.59 which seems to be consistent with (nearly) flat bottom seen in the profile. We note that, \of\ profile is unusually broad which could possibly due to unknown contamination. Therefore, the actual $f_c$ for \of\ could be even less. Because of the weak line strength we do not attempt to estimate $f_c$ for \nani, instead we use \neo\ covering fraction for fitting. We  note that unlike \neo, profiles of \nani\ are not saturated.    

\begin{figure} 
\centerline{
\vbox{
\centerline{\hbox{ 
\includegraphics[height=4.4cm,width=8.4cm,angle=00]{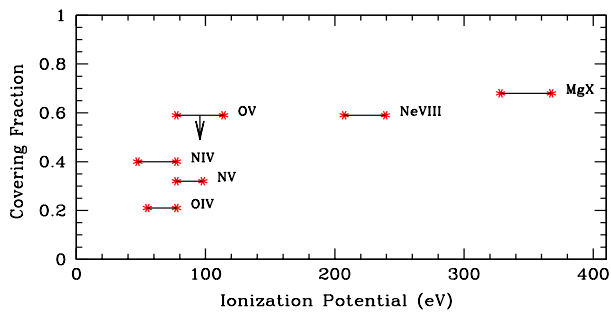} 
}}
}}
\caption{Covering fractions of different species detected in \zabs\ = 1.02854 
towards PG~1206$+$459 as a function of ionization potential. The energy range 
between the creation and destruction ionization potentials of a given species 
are shown by the two stars connected by a solid line.      
} 
\label{IP_fc} 
\end{figure}

In Fig.~\ref{IP_fc}, we have plotted the covering fractions ($f_c$) of different species detected in this system as a function of ionization potentials. It is clear from the figure that we have two sets of covering fractions for this system. The species with high ionization potentials (i.e. \neo, \mgx) are showing covering fraction $f_c \gtrsim 0.6$, whereas, the low ionization species (i.e. \ofo, \nfo, \nf) show $f_c \lesssim 0.4$. Ionization potential dependent covering fraction have already been reported by \citet[]{Telfer98,Muzahid12b}, where, the idea of multiphase structure of the absorbing gas has been put forward, with different species having different projected area. In view of this, the covering fraction of \nani\ should be similar to that of \neo, as they have ionization potentials of the same order. This also justifies our use of \neo\ covering fraction for the fitting of \nani\ doublets. Such an assumption indeed gives good fit to \nani\ doublets.    

%
%

\begin{table}
\caption{Partial coverage corrected Voigt profile fit parameters for the absorber at \zabs\ = 1.02854 towards PG~1206$+$459.}  
\begin{tabular}{ccccc} 
\hline 
\hline 
 $v_{\rm ej}$(\kms) & Ion & $b$(\kms) & log~$N$(cm$^{-2}$) & $f_c$ \\   
  (1)    &    (2)               & (3)   &  (4)       & (5)         \\ 
\hline 
\hline 
 $-$19290  &   \mgx\   &   65 $\pm$ 4 &   15.31 $\pm$ 0.06 & 0.68 ($db$) \\
 $-$19314  &  \nani\   &   93 $\pm$12 &   14.93 $\pm$ 0.05 & 0.59 ($aa$) \\ 
 $-$19285  &   \neo\   &   83 $\pm$ 4 &   $>$15.90 $\pm$ 0.06 & 0.59 ($fb$) \\
 $-$19285  &  \arei\   &   83 $\pm$ 0 & $<$14.19 $\pm$ 0.02 & 0.59 ($aa$) \\ 
 $-$19285  &    \of\   &  141 $\pm$ 5 &   $>$14.92 $\pm$ 0.03 & 0.59 ($fb$) \\
 $-$19339  &    \nf\   &   72 $\pm$18 &   $>$15.38 $\pm$ 0.40 & 0.32 ($fb$) \\
 $-$19285  &   \ofo\   &   57 $\pm$11 &   $>$15.85 $\pm$ 0.23 & 0.21 ($fb$) \\
 $-$19272  &   \nfo\   &   94 $\pm$17 &   $>$14.77 $\pm$ 0.18 & 0.40 ($fb$) \\ 
 $-$19285  &    \hi\   &   83 $\pm$ 0 &   $\sim$13.78      & 0.59 ($aa$) \\ 
 $-$19285  &    \hi\   &   83 $\pm$ 0 &   $\sim$14.42      & 0.30 ($aa$) \\ 
\hline                                                                           
 $-$19163  &   \mgx\  &   88 $\pm$ 9 &   15.28 $\pm$ 0.06 & 0.68 ($db$) \\
 $-$19150  &   \nani\ &   70 $\pm$11 &   14.68 $\pm$ 0.08 & 0.59 ($aa$) \\ 
 $-$19140  &    \neo\ &   68 $\pm$ 5 &   $>$15.69 $\pm$ 0.11 & 0.59 ($fb$) \\
 $-$19140  &   \arei\ &   68 $\pm$ 0 & $<$13.66 $\pm$ 0.07 & 0.59 ($aa$) \\ 
 $-$19140  &     \of\ &  113 $\pm$11 &   $>$14.97 $\pm$ 0.05 & 0.59 ($fb$) \\
 $-$19202  &   \nf\   &   70 $\pm$26 &   $>$15.21 $\pm$ 0.38 & 0.32 ($fb$) \\
 $-$19140  &    \ofo\ &   53 $\pm$12 &   $>$15.40 $\pm$ 0.13 & 0.21 ($fb$) \\
 $-$19140  &    \nfo\ &   46 $\pm$15 &   $>$14.28 $\pm$ 0.36 & 0.40 ($fb$) \\ 
 $-$19140  &    \hi\  &   68 $\pm$ 0 &   $\sim$13.61      & 0.59 ($aa$) \\
 $-$19140  &    \hi\  &   68 $\pm$ 0 &   $\sim$14.01      & 0.30 ($aa$) \\
\hline 
\hline 
\end{tabular}
\vfill ~\\ 
Note -- Listed errors on all the quantities in this paper only include the statistical 
errors. For $v_{\rm ej}$ the COS calibration uncertainty is $\sim \pm$~10 \kms.  In addition, 
the uncertainty in the COS LSF introduces errors of at least 1 to 3 \kms\ in these profile 
fit line widths. Zero error implies that the parameter was tied/fixed during fitting. 
Covering fraction, $f_c$ used to estimate the column density is given in column~5. Method used 
to compute $f_c$ is mentioned in parenthesis. ``$fb$"-- from flat bottom profile, ``$db$"-- from 
doublets, ``$aa$"-- physically motivated assumed value. Note that the column density estimated 
from the flat bottom profile (i.e. ``$fb$") should be taken as lower limit.         
\label{tab_pg1206}  
\end{table}  

Parameters estimated through Voigt profile fitting after correcting for partial coverage are given in Table~\ref{tab_pg1206}. We treat column densities of all the species as lower limits in the case of ions showing flat bottom profiles. We note that, both the members of \arei\ doublets are partially blended by unknown contaminants which made the covering fraction estimation impossible. However, since the ionization potentials (creation$+$destruction) of \arei\ are comparable to those of \of, we take $f_c = 0.59$. We note that, because of blend $N(\arei)$  should be taken as upper limit.  

At the expected position of \os, some absorption is seen in the low resolution $HST/$FOS G190H spectrum. However, due to severe blending in both members of the doublets, we do not attempt to estimate the covering fraction. Estimated conservative upper limit on \os\ column density is log~$N(\os)$ [cm$^{-2}$] $<$ 14.80, assuming $f_c$ = 0.59. In addition, we do not detect any clear signature of \lyb\ absorption in G190H spectrum. Weak \lya\ absorption line, seen in STIS/E230M spectrum, is fitted with two different values of covering fractions (i.e. $f_c$ = 0.59 and 0.30; see Table~\ref{tab_pg1206}), in order to estimate the maximum \hi\ content associated with the high and low ionization phases. However, in both the phases $N(\hi)$ found to be $< 10^{14.5}$ cm$^{-2}$. All these suggest a very little neutral hydrogen content in this absorber. 

\subsection{\zabs = 1.21534 towards PG~1338$+$416}  
\label{sec_discript_PG1338_1.21534}  

\begin{figure*} 
\begin{sideways}
\begin{minipage}{24cm} 
\centerline{\vbox{
\centerline{\hbox{
\includegraphics[height=11.0cm,width=8.0cm,angle=00]{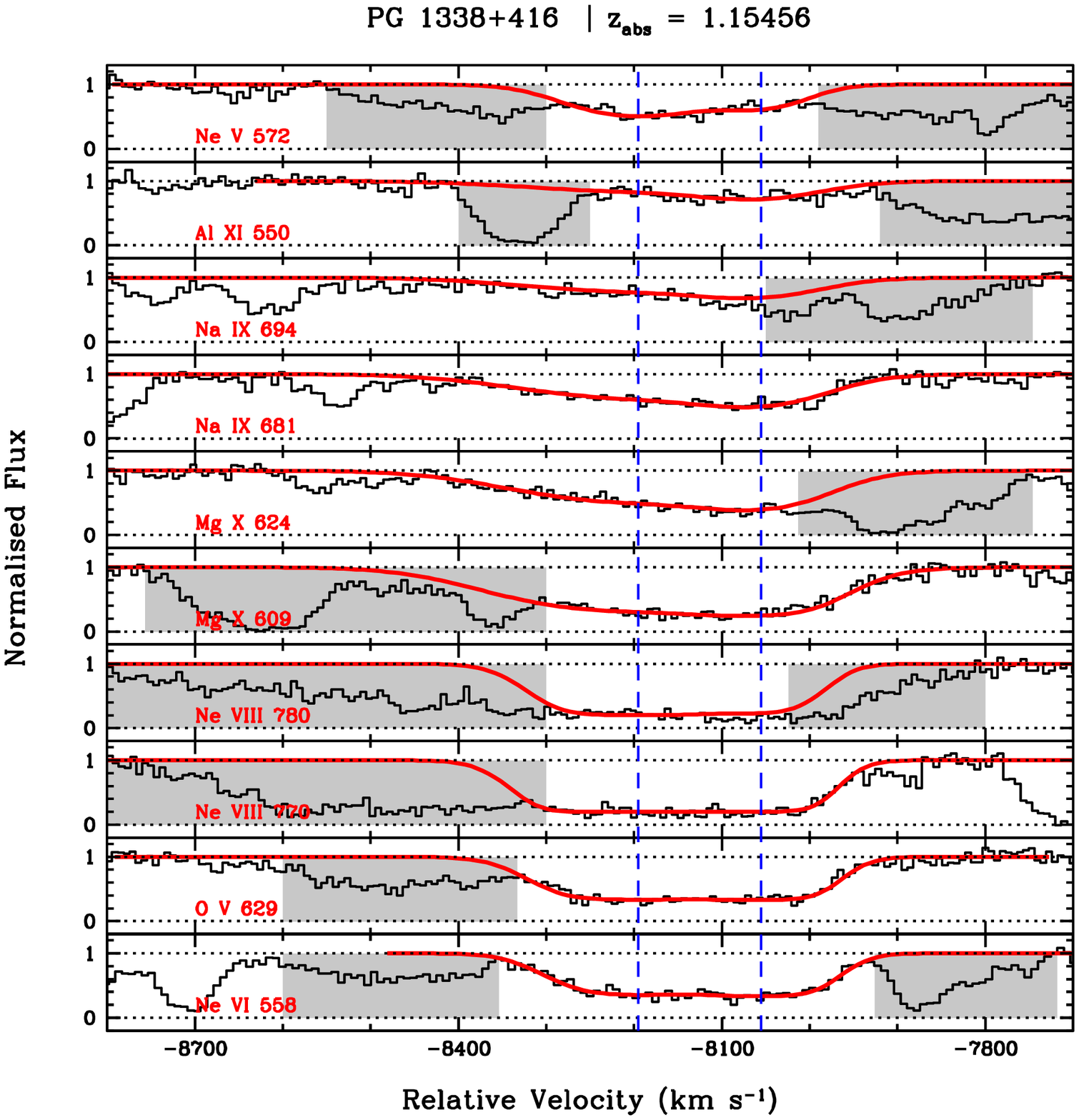} 
\includegraphics[height=11.0cm,width=8.0cm,angle=00]{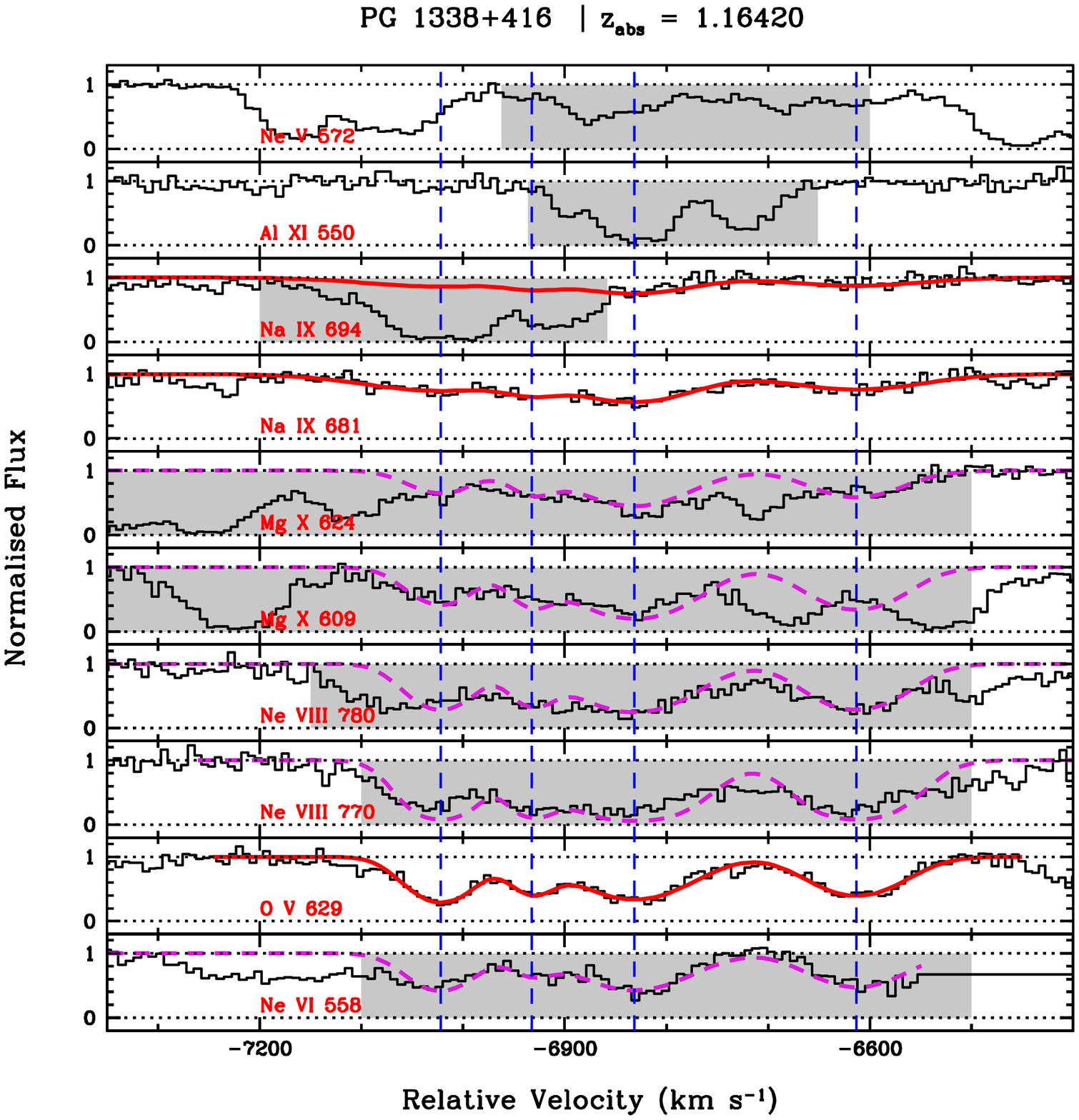} 
\includegraphics[height=11.0cm,width=8.0cm,angle=00]{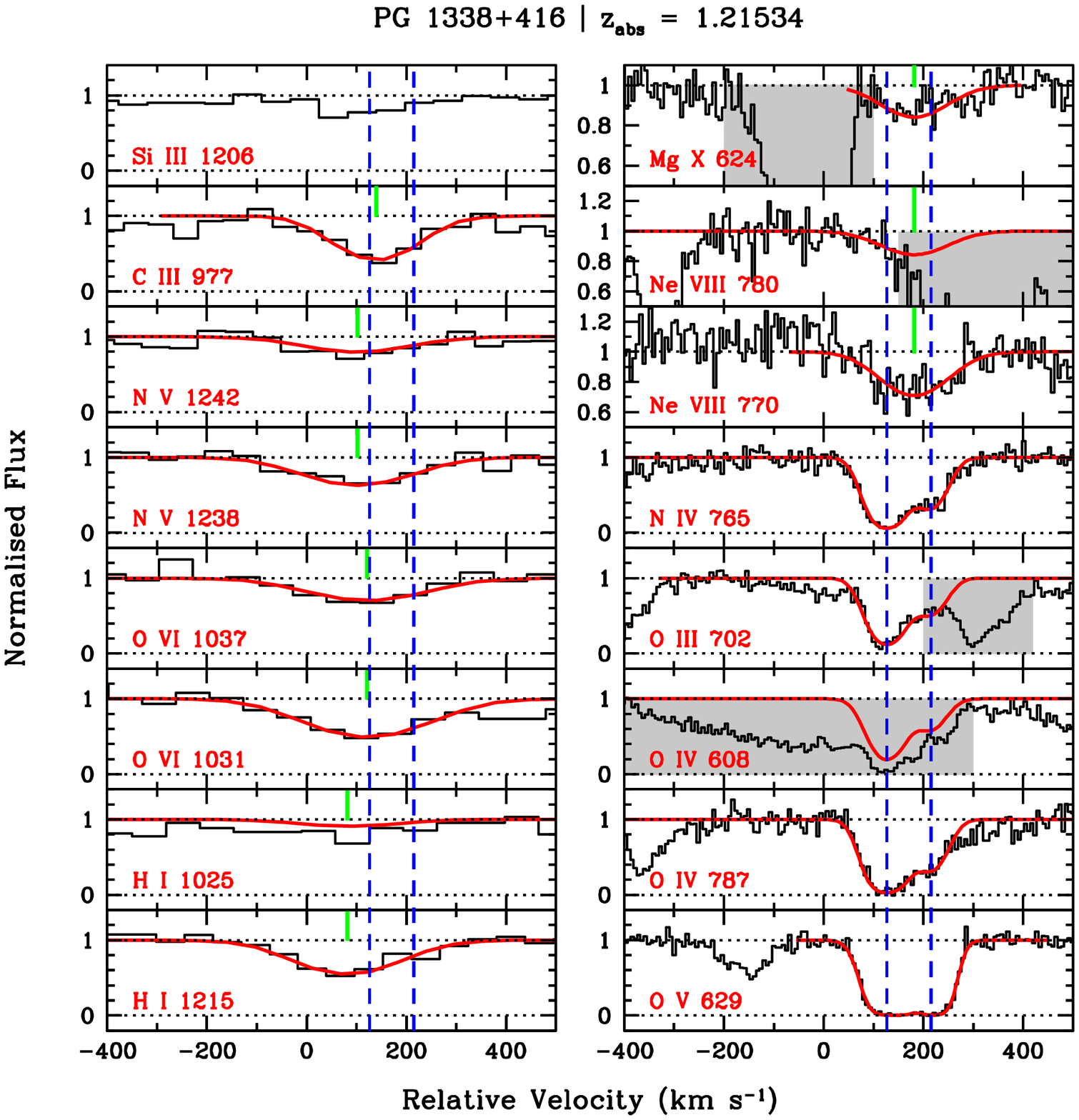} 
}}
}} 
\caption{Velocity plot of the three associated \neo\ systems detected towards PG~1338$+$416.  
The zero velocity corresponds to the emission redshift (\zem\ = 1.214) of the QSO. The 
smooth curves overplotted on top of the data are the best fitting Voigt profiles after 
correcting for the partial coverage whenever needed. The shaded regions mark the contamination 
due to unrelated absorption to the system of interest. The vertical dashed lines mark the 
positions of individual Voigt profile components. 
{\sl Left :} The system at \zabs\ = 1.15456. {\sl Middle :} The system at \zabs\ = 1.16420. 
The smooth dashed curves are not fit to the data, but the synthetic profiles corresponding 
to the maximum allowed column density assuming complete coverage (see text). 
{\sl Right :} The system at \zabs\ = 1.21534. Apart from \ct\ all other species plotted in 
the left hand sub-panel are from  FOS/G270H spectrum whereas \ct\ is from FOS/G190H spectrum.  
All the species plotted in the right hand sub-panel are from COS spectrum. Two components are 
clearly seen in \oth, \ofo, \nfo\ and \of\ absorption (shown by two vertical dashed lines).  
In all other cases single component is needed as shown by (green) solid tick.   
} 
\label{vp_pg1338} 
\end{minipage}
\end{sideways} 
\end{figure*}

The ejection velocity of this system is $v_{\rm ej} \sim +181$~\kms\ suggesting \zabs~$>$~\zem. This absorber (see the rightmost panel of Fig.~\ref{vp_pg1338}) is primarily detected through the presence of \os\ doublets in FOS/G270H spectrum and subsequently confirmed with various other low ionization species (e.g. \oth, \nfo, \ofo, \of\ etc.) in COS spectrum. Weak absorption from high ionization species like \mgx\ and \neo\ are also detected. However, \neo~$\lambda780$ profile is blended with strong \lyb\ absorption from \zabs\ = 0.6863 system. The \mgx\ $\lambda 609$ line is heavily blended, possibly with low redshift \lya\ line and hence not shown in the figure. The non-detection of \nani\ $\lambda 681$ is consistent with log~$N(\nani)[{\rm cm^{-2}}] < 14.18$ at 3$\sigma$ confidence level. \ofo~$\lambda608$ line is severely blended with \mgx~$\lambda624$ line from \zabs\ = 1.15456. \oth~$\lambda702$ line is partially blended with unknown contaminants.      
The uncontaminated low ionization species (i.e. \nfo~$\lambda765$, \ofo~$\lambda787$) clearly show multicomponent structure. In both cases, at least two Voigt profile components (shown by vertical dashed lines) are required to get best fitted $\chi^{2}$ close to 1. Unlike low ionization species, \neo$\lambda$770 absorption shows smooth and/or broad profile which is well fitted by a single component. Due to poor spectral resolution, all the ions detected in FOS can be fitted with a single component.  

We do not find a clear signature of partial coverage in any line. For example, \nf\ and \os\ doublets are well fitted with $f_c$ = 1.0 and do not show non-zero flat bottom profiles. The \lya\ and (weak) \lyb\ absorption are also consistent with complete coverage of the background source by the absorber. The Voigt profile fit parameters for this absorber are given in Table~\ref{tab3_pg1338}. We would like to mention here that, because of blending in \oth\ line and saturation in \of\ line, $N(\oth)$ and $N(\of)$ should be taken as upper and lower limits respectively. We will use these bounds in section~\ref{sec_phot_model2_pg1338}, where we discuss the photoionization modelling of this system. In passing, we note that \hi\ and \os\ line centroids are offset by $\sim$40 \kms. This could be  a signature of multiphase gas. We also note that, \hi\ and \ct\ line centroids are offset by $\sim$60 \kms. However, as they are detected in spectra taken with two different instruments (i.e. FOS G270H and G190H), such an offset could also be attributed to the systematic uncertainties.

\begin{table}
\begin{center}
\caption{Voigt profile fit parameters for the absorber at \zabs\ = 1.21534 towards 
PG~1338$+$416 using $f_c = 1$ for all the species.} 
\begin{tabular}{rccc} 
\hline 
 $v_{\rm ej}$(\kms) & Ion & $b$(\kms) & log~$N$(cm$^{-2}$) \\   
  (1)    &    (2)               & (3)   &  (4)                 \\ 
\hline 
\hline 
  $+$81  &   \hi\  &  137 $\pm$14 &  14.04 $\pm$ 0.04  \\
 $+$101  &   \nf\  &  144 $\pm$15 &  14.36 $\pm$ 0.04  \\
 $+$121  &   \os\  &  158 $\pm$49 &  14.74 $\pm$ 0.12  \\
 $+$126  &  \nfo\  &   44 $\pm$ 2 &  14.25 $\pm$ 0.03  \\
 $+$126  &  \ofo\  &   44 $\pm$ 2 &  15.08 $\pm$ 0.03  \\
 $+$126  &  \oth   &   44 $\pm$ 0 &  14.82 $\pm$ 0.01  \\
 $+$136  &   \of\  &   45 $\pm$ 3 &  14.78 $\pm$ 0.05  \\
 $+$139  &   \ct\  &  105 $\pm$17 &  13.91 $\pm$ 0.06  \\
 $+$181  &  \neo\  &   91 $\pm$12 &  14.42 $\pm$ 0.05  \\
 $+$181  &  \mgx\  &   91 $\pm$ 0 &  14.62 $\pm$ 0.06  \\
 $+$214  &  \nfo\  &   40 $\pm$ 5 &  13.80 $\pm$ 0.06  \\
 $+$214  &  \ofo\  &   40 $\pm$ 5 &  14.53 $\pm$ 0.07  \\
 $+$214  &  \oth\  &   40 $\pm$ 0 &  14.26 $\pm$ 0.03  \\
 $+$223  &   \of\  &   32 $\pm$ 3 &  14.55 $\pm$ 0.06  \\
\hline 
\hline 
\end{tabular}
\label{tab3_pg1338}  
\end{center}
\end{table}  

\subsection{\zabs = 1.16420 towards PG~1338$+$416}   
\label{sec_discript_PG1338_1.16420}

The ejection velocity of this system is $v_{\rm ej} \sim -6818$~\kms. The velocity plot of this systems is shown in the middle panel of Fig.~\ref{vp_pg1338}. Apart from \of\ and \nani~$\lambda 681$, all other detected ions in this system show complex blend in their profiles. The overall similarity in profiles of various ions clearly assures their presence. We do not find any other contamination in \nani~$\lambda 681$ absorption and it shows very similar profile like \of. Therefore, we believe \nani\ detection is robust, although the blue wing of \nani~$\lambda 694$ line is severely blended. \nesi~$\lambda 558$ absorption falls near the Galactic Ly$\alpha$ absorption and hence the continuum around this absorption is not well constrained. Since \of\ is singlet transition and \nani~$\lambda 694$ is blended, we did not estimate covering fraction for any of these ions. We assume $f_c = 1$ to get a lower limit on column densities. 
At least four Voigt profile components are required to fit the unblended \of\ and \nani~$\lambda 681$ profiles. The \nani\ to \of\ column density ratio in all four components are consistent within factor $\sim$~2 (e.g. log~$N(\nani)/N(\of)$ = 0.27$\pm$0.21). Because of contamination in the case of \mgx\ and \neo\ lines and poorly constrained continuum in the case of \nesi~$\lambda 558$ line we do not perform Voigt profile fitting for these absorption. Instead, we check the consistency of synthetic profiles generated using the component structure and the $b$-parameters similar to \of\ line, assuming $f_c = 1.0$. The synthetic profiles are shown in smooth dashed curves on top of data, in the middle panel of Fig.~\ref{vp_pg1338}. The highest optical depth pixels in \neo\ doublets are roughly consistent with $f_c \gtrsim 0.8$. The line measurements for this system are presented in Table~\ref{tab2_pg1338}.   
The low resolution FOS/G190H spectrum shows  absorption in the expected position of \os. However, contamination of \os\ lines from \zabs\ = 1.15456 absorber do not allow any reliable column density estimation. \lya\ and \lyb\ absorption from this absorber are covered by the G270H and G190H spectra respectively. However, \lyb\ is found to be stronger than \lya, suggesting a possible contamination in \lyb. \lya, on the other hand, is contaminated with Galactic Fe~{\sc ii} lines. Therefore, we do not present any measurement for \hi\ in Table~\ref{tab2_pg1338}. Due to poorly constrained $f_c$, the column density measurements are highly uncertain and hence we do not discuss the ionization modelling for this system, in spite of the presence of \nani.  

\subsection{\zabs = 1.15456 towards PG~1338$+$416}  
\label{sec_discript_PG1338_1.15456}  

\begin{table}
\begin{center} 
\caption{Voigt profile fit parameters for \zabs\ = 1.16420 towards PG~1338$+$416 
assuming $f_c = 1$ for all the species.}  
\begin{tabular}{cccc} 
\hline 
 $v_{\rm ej}$(\kms) & Ion & $b$(\kms) & log~$N$(cm$^{-2}$)   \\   
  (1)    &    (2)               & (3)   &  (4)                   \\ 
\hline 
\hline 
$-7022$    & \nani\   &   100$\pm$ 13 &     $>$ 14.50 $\pm$ 0.05  \\
           &    \of\  &   40 $\pm$  3 &     $>$ 14.03 $\pm$ 0.02  \\
           &    \mgx\ &   40          &           $\sim$   14.67   \\
           &    \neo\ &   40          &           $\sim$   14.94   \\
           &   \nesi\ &   40          &           $\sim$   14.68   \\ 
\hline 
$-6932$	   & \nani\   &   34 $\pm$ 11 &      $>$ 13.91 $\pm$ 0.15  \\  	
           &    \of\  &	  27 $\pm$  4 &      $>$ 13.68 $\pm$ 0.07  \\      
           &    \mgx\ &	  27          &           $\sim$   14.67   \\      
	   &    \neo\ &	  27          &           $\sim$   14.50   \\      
           &   \nesi\ &	  27          &           $\sim$   14.14   \\      
\hline 
$-6832$	   & \nani\   &  74 $\pm$ 7   &      $>$ 14.65 $\pm$ 0.04  \\ 	 
           &    \of\  &  66 $\pm$ 4   &      $>$ 14.17 $\pm$ 0.02  \\ 
           &    \mgx\ &  66           & $\sim$   15.20   \\ 
	   &    \neo\ &  66           & $\sim$   15.14   \\ 
           &   \nesi\ &  66           & $\sim$   14.87   \\  
\hline 
$-6613$	   & \nani\   &  86 $\pm$ 14  &     $>$  14.40 $\pm$ 0.06  \\  
           &    \of\  &  57 $\pm$  3  &     $>$  14.04 $\pm$ 0.02  \\ 
           &    \mgx\ &  57           & $\sim$   15.09   \\ 
	   &    \neo\ &  57           & $\sim$   14.91   \\ 
           &   \nesi\ &  57           & $\sim$   14.75   \\ 
\hline 
\hline 
\end{tabular}
\label{tab2_pg1338}  
\end{center} 
\end{table}  

\begin{figure} 
\centerline{
\vbox{
\centerline{\hbox{ 
\includegraphics[height=8.4cm,width=8.8cm,angle=00]{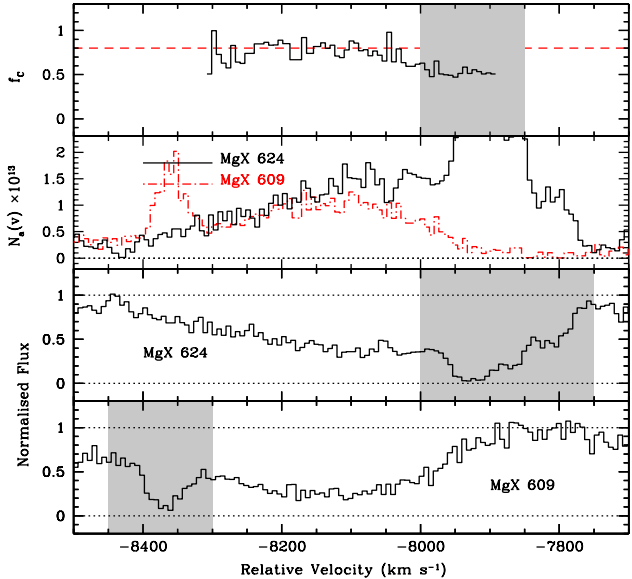} 
}}
}}
\caption{Profiles of \mgx\ doublet for the system at \zabs\ = 1.15456 towards PG~1338$+$416 
are shown in the two bottom panels. Corresponding apparent column density distributions 
[in units of 10$^{13}$ cm$^{-2}$ (km s$^{-1}$)$^{-1}$] are shown in the second panel from 
the top. The covering fraction distribution is shown in the topmost panel. The dashed line 
indicates the median value of covering fraction $f_c = 0.8 \pm 0.1$, as measured in the core 
pixels. The shaded regions show the velocity range affected by unrelated absorption.
} 
\label{covf_pg1338} 
\end{figure}

The ejection velocity of this system is $v_{\rm ej} \sim -8156$~\kms, with \neo\ absorption spread over $\sim 340$~\kms. The profiles of different species originating from this system are plotted as a function of outflow velocity in the leftmost panel of Fig.~\ref{vp_pg1338}. This is the highest \mgx\ column density system in our sample. The core pixels of \mgx\ doublets are free from any blend and clearly show broad multicomponent structure with at least two components contributing to the absorption. We will refer to the highest velocity component as component-1 and the other as component-2\ in subsequent discussions regarding this system (e.g. in section~\ref{sec_phot_model1_pg1338}). There is only a mild contamination from Ly$\gamma$ of \zabs\ = 0.3488 absorber in the blue wing of the \mgx~$\lambda609$ line as shown by shaded region. The red wing of the \mgx~$\lambda624$, on the other hand, is blended with \ofo~$\lambda608$ transitions from another associated absorber (i.e. \zabs\ = 1.21534) along this sight line. We use the uncontaminated core pixels (i.e., between $-8350 < v~(\rm km s^{-1}) < -8150$) of \mgx\ doublets and estimate the covering fraction $f_c = 0.8\pm0.1$ (see Fig.~\ref{covf_pg1338}). \nani~$\lambda681$ line is completely free from any contamination and shows remarkable similarity with \mgx\ profiles. This possibly means \mgx\ and \nani\ are originating from the same phase of the absorbing gas. The red wing of \nani~$\lambda694$ line, however, is blended by Ly$-9$ transition from a previously known DLA at \zabs\ = 0.6214 \citep[]{Rao06}. Therefore we use covering fraction for \nani\ similar to that of \mgx. We note that such an assumption gives remarkably good fit to \nani\ doublets. 
%

Both transitions of \neo\ doublet show very strong, albeit blended, absorption with flat bottom profiles consistent with $f_c = 0.8$. The strong uncontaminated \of\ and \nesi\ lines also show flat bottom profiles. The covering fraction in these two cases, as calculated from the flat bottom, are very similar and lower (i.e. $f_c$ = 0.67) than that of very highly ionized species (i.e. \neo, \mgx). Clearly, like the previous case (i.e. \zabs\ = 1.02854 towards PG~1206$+$459), here also we find two sets of covering fraction for the detected species suggesting ionization potential dependent phase separation of the absorbing gas. \nefi\ line seen in this absorber is unsaturated and shows two possible velocity components. However, due to severe blending in both the wings of \nefi\ absorption, we only estimate the upper limit on $N(\nefi)$ assuming $f_c$ and $b$-parameters similar to those of \of\ line. In section~\ref{sec_phot_model}, we will show that, under photoionization equilibrium \of\ and \nefi\ trace each other for the whole range of ionization parameters. Therefore, using the \of\ covering fraction for \nefi\ is legitimate.  We also estimate upper limits on the weak absorption seen in the expected position of \alel~$\lambda550$ transition assuming $f_c$ and $b$-parameters similar to those of \mgx, as they have ionization potentials of similar order. However, as both the wings of \alel~$\lambda550$ line is blended the measured column density is merely a upper limit. The other member of \alel\ doublet with $\lambda_{\rm rest} = 568$~\AA, is severely affected by the Galactic \lya\ absorption and complex blend. In addition, we do not detect any clear signature of \lya\ absorption, in the FOS/G270H spectrum. Some absorption is seen in the expected positions of \os\ doublets in the FOS/G190H spectrum. However, the contamination of \os\ lines from \zabs\ = 1.16420 absorber and the poor data quality prevent us from any reliable column density estimations.   
The partial coverage corrected Voigt profile fit parameters are given in Table~\ref{tab1_pg1338}. In the case of non-detections (i.e. \hi, \ofo\ and \arei), we present 3$\sigma$ upper limits on column densities as estimated from the error in the continuum. 

\section{Ionization models} 
\label{sec_phot_model} 

\begin{figure} 
\centerline{
\vbox{
\centerline{\hbox{ 
\includegraphics[height=8.4cm,width=8.4cm,angle=00]{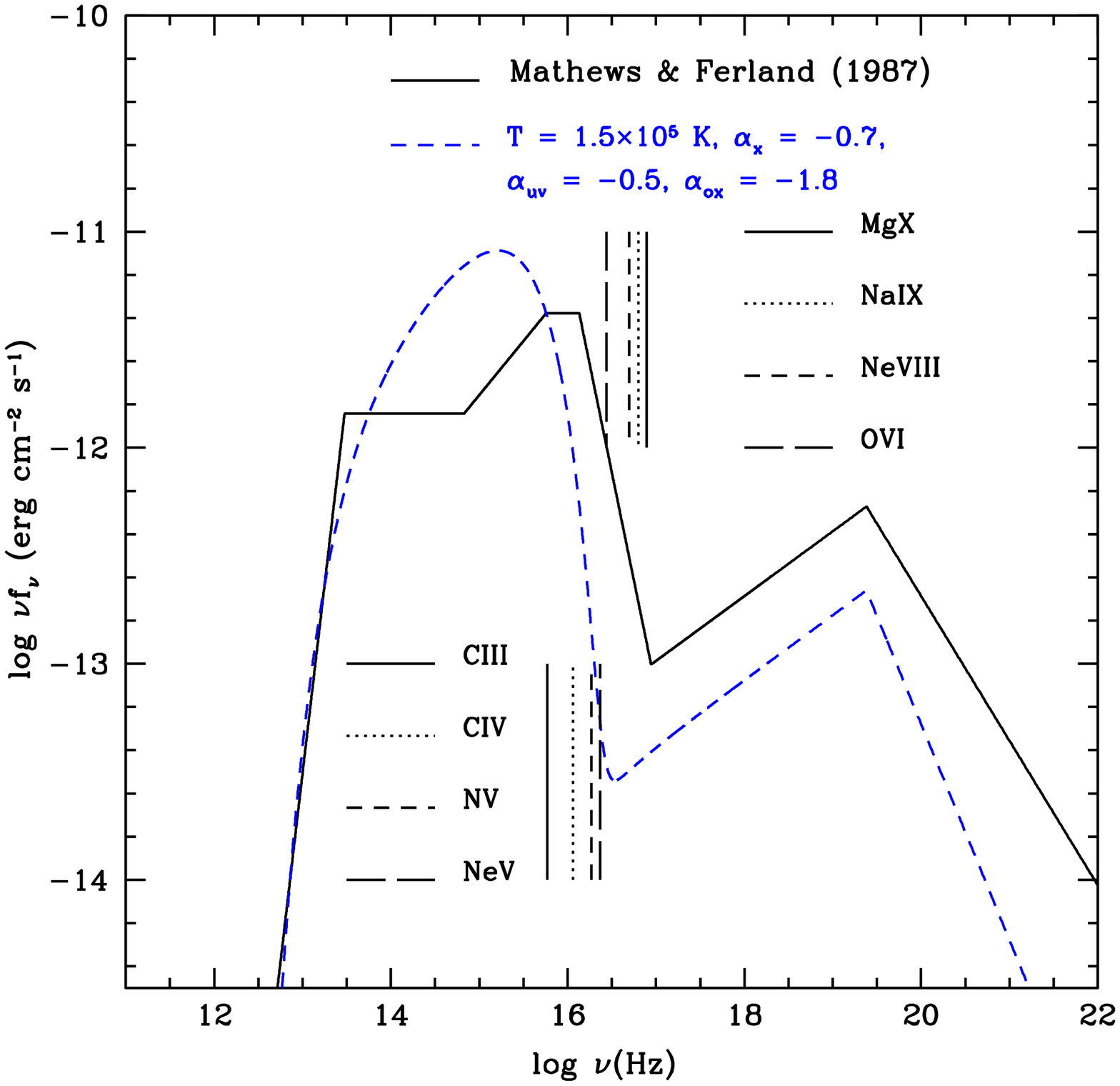} 
}}
}}
\caption{Typical shapes of spectral energy distributions of an AGN with arbitrary 
normalization. The solid line gives the spectrum by \citet{Mathews87} whereas the 
dotted curve is generated assuming a blackbody with temperature 
$T_{\rm BB} \sim 1.5\times10^{5}$~K and power laws with typical slopes, 
$\alpha_{\rm uv} = -0.5$, $\alpha_{\rm x} = -0.7$ and $\alpha_{\rm ox} = -1.8$ 
(see text). The vertical lines with different line styles mark the frequency 
corresponding to the ionization potential of the species mentioned in the plot.  
} 
\label{sed_mf} 
\end{figure} 
%
 
\begin{table}
\caption{Partial coverage corrected Voigt profile fit parameters for \zabs\ = 
1.15456 towards PG~1338$+$416.}  
\begin{tabular}{ccccc} 
\hline 
 $v_{\rm ej}$(\kms) & Ion & $b$(\kms) & log~$N$(cm$^{-2}$) & $f_c^{a}$ \\   
  (1)    &    (2)               & (3)   &  (4)       & (5)         \\  
\hline 
\hline 
$-$8195   & \nani\  &  165 $\pm$ 12   & 15.05 $\pm$ 0.03 & 0.80 ($aa$) \\ 
          & \mgx\   &  165 $\pm$  7   & 15.62 $\pm$ 0.02 & 0.80 ($db$) \\ 
	  & \alel\  &        165      & $\le$14.77 $\pm$ 0.07 & 0.80 ($aa$) \\  
          & \neo\   &   89 $\pm$ 35   & $>$15.94 $\pm$ 0.41 & 0.80 ($fb$) \\ 
          & \of\    &   91 $\pm$  6   & $>$14.97 $\pm$ 0.06 & 0.67 ($fb$) \\ 
          & \nesi\  &   87 $\pm$  5   & $>$15.57 $\pm$ 0.04 & 0.67 ($fb$) \\ 
	  & \ofo\   &         91      & $\le$ 13.69      & 0.67 ($aa$) \\  
	  & \nefi\  &         91      & $\le$ 15.26      & 0.67 ($aa$) \\ 
	  & \nefo\  &         91      & $\le$ 13.66      & 0.67 ($aa$) \\ 
          & \arei\  &         91      & $\le$ 13.94      & 0.67 ($aa$) \\ 
	  & \hi\    &        165      & $\le$ 13.64      & 0.67 ($aa$) \\   
	  & \hi\    &        165      & $\le$ 13.56      & 0.80 ($aa$) \\   
\hline                                                   
$-$8055   & \nani\  &   90 $\pm$  8   & 14.82 $\pm$ 0.04 & 0.80 ($aa$) \\  
          & \mgx\   &   86 $\pm$  5   & 15.33 $\pm$ 0.03 & 0.80 ($db$) \\  
	  & \alel\  &        86       & $\le$14.66 $\pm$ 0.06 & 0.80 ($aa$) \\  
          & \neo\   &   62 $\pm$  6   & $>$15.45 $\pm$ 0.12 & 0.80 ($fb$) \\    
          & \of\    &   65 $\pm$  5   & $>$14.85 $\pm$ 0.08 & 0.67 ($fb$) \\ 
          & \nesi\  &   66 $\pm$  5   & $>$15.60 $\pm$ 0.07 & 0.67 ($fb$) \\  
	  & \ofo\   &         65      & $\le$ 13.90      & 0.67 ($aa$) \\    
	  & \nefi\  &         65      & $\le$ 14.87      & 0.67 ($aa$) \\ 
	  & \nefo\  &         65      & $\le$ 13.10      & 0.67 ($aa$) \\ 
	  & \arei\  &         65      & $\le$ 13.30      & 0.67 ($aa$) \\ 
	  & \hi\    &         91      & $\le$ 13.50      & 0.67 ($aa$) \\   
	  & \hi\    &         91      & $\le$ 13.42      & 0.80 ($aa$) \\   
\hline  
\hline 
\end{tabular}
\vfill ~\\ 
Table Note -- $^{a}$Same as Table~\ref{tab_pg1206}
\label{tab1_pg1338}  
\end{table}  
%

%
\begin{figure*} 
\centerline{
\vbox{
\centerline{\hbox{ 
\includegraphics[height=6.4cm,width=8.4cm,angle=00]{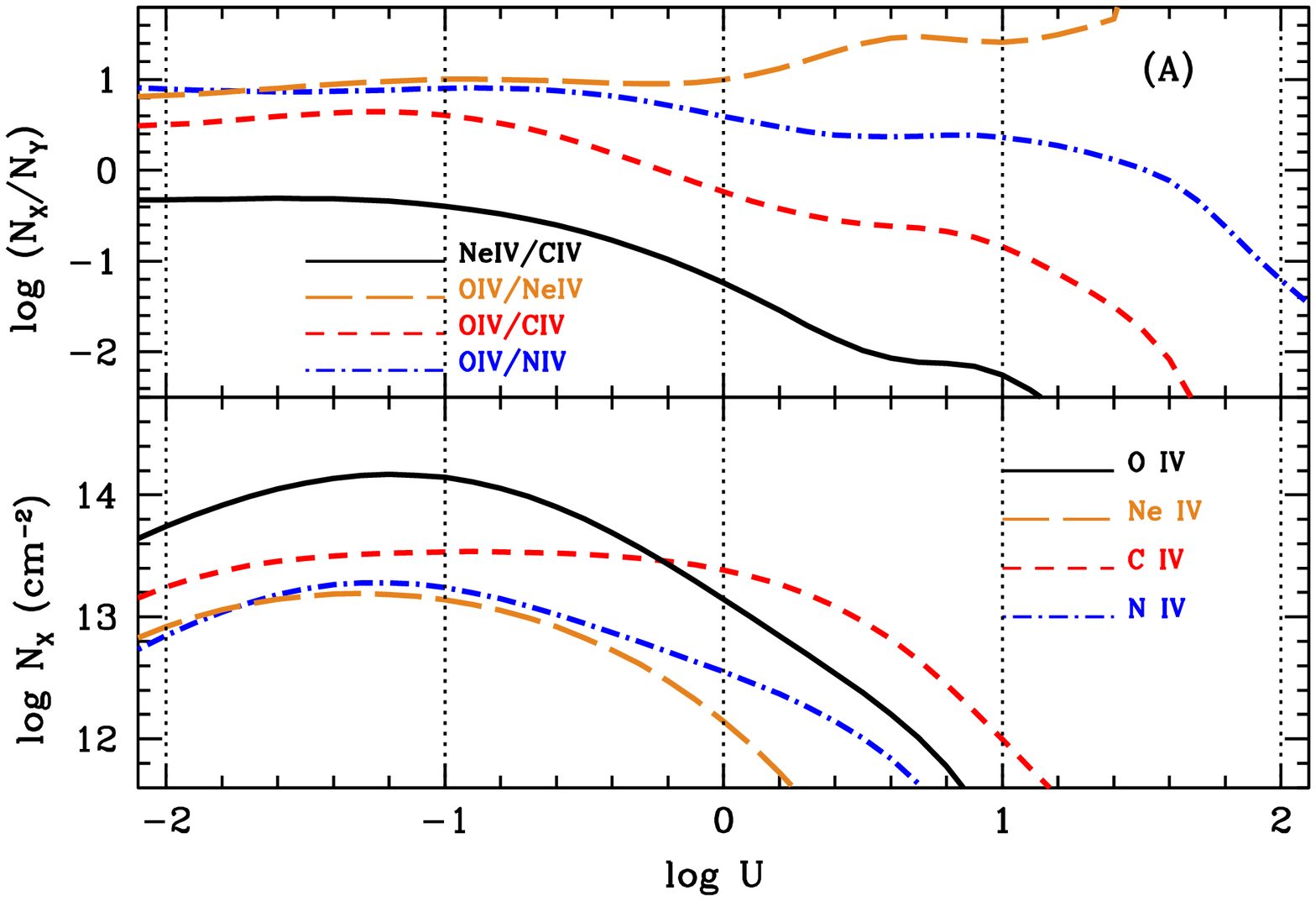} 
\includegraphics[height=6.4cm,width=8.4cm,angle=00]{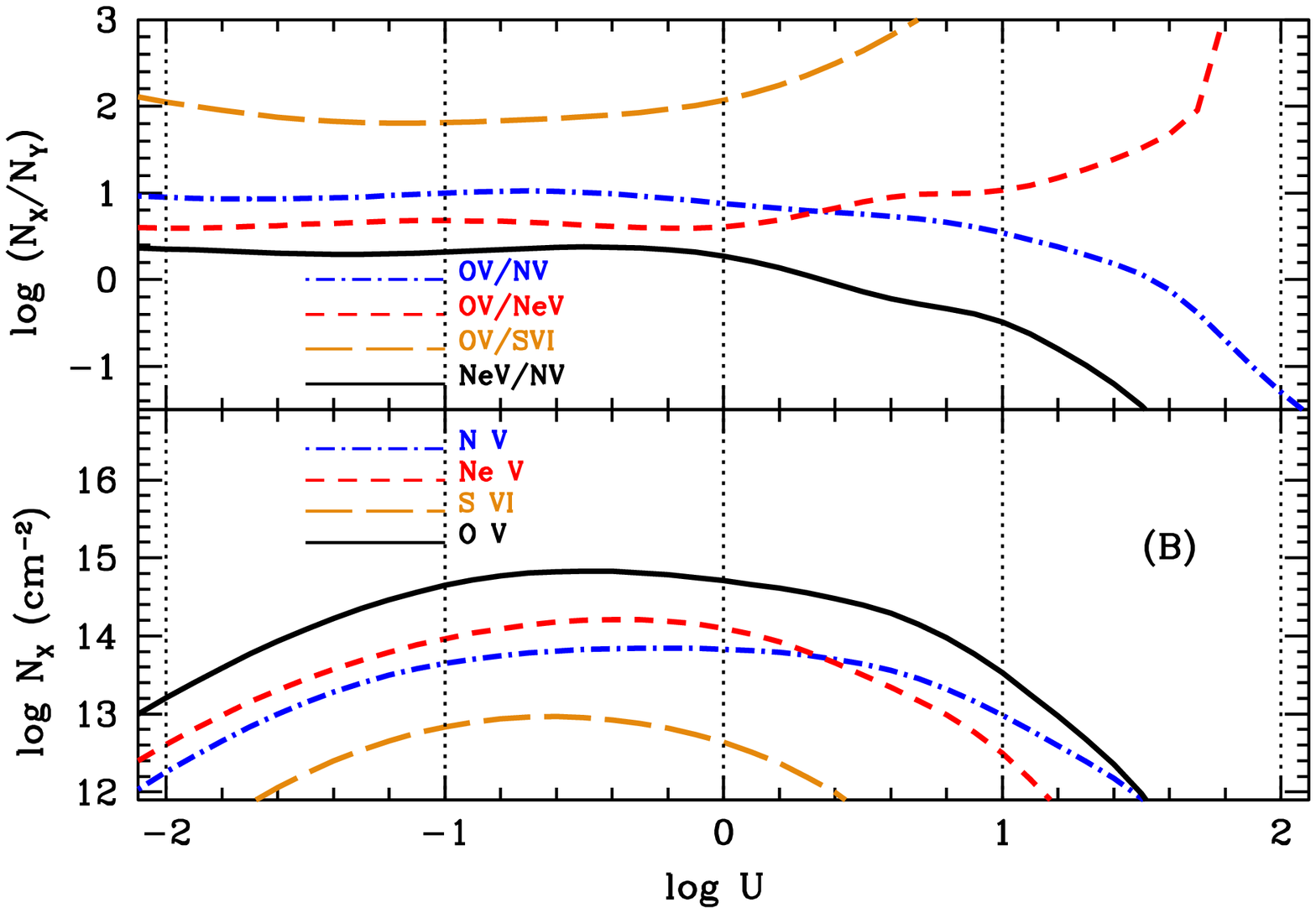} 
}}
\centerline{\hbox{ 
\includegraphics[height=6.4cm,width=8.4cm,angle=00]{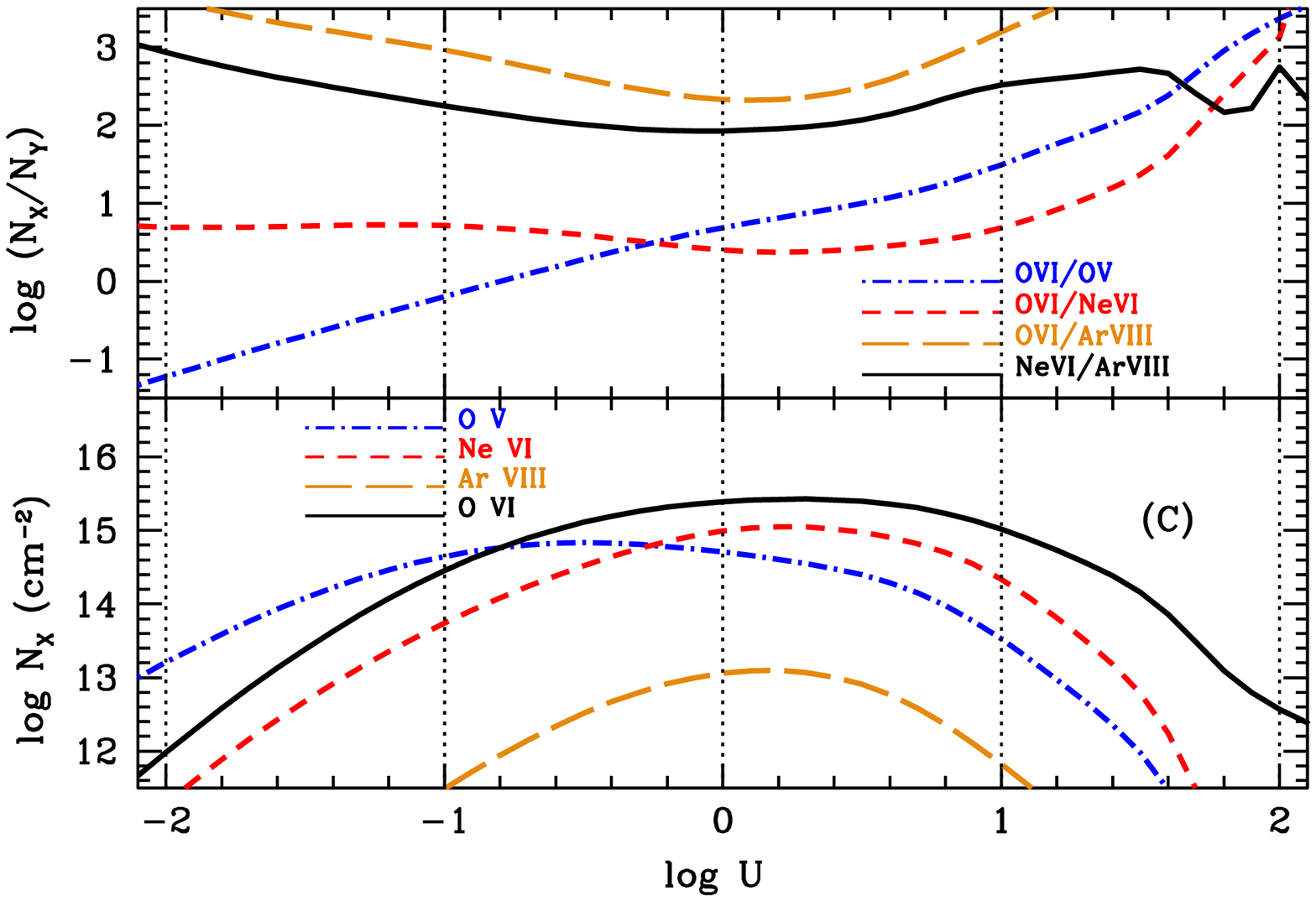} 
\includegraphics[height=6.4cm,width=8.4cm,angle=00]{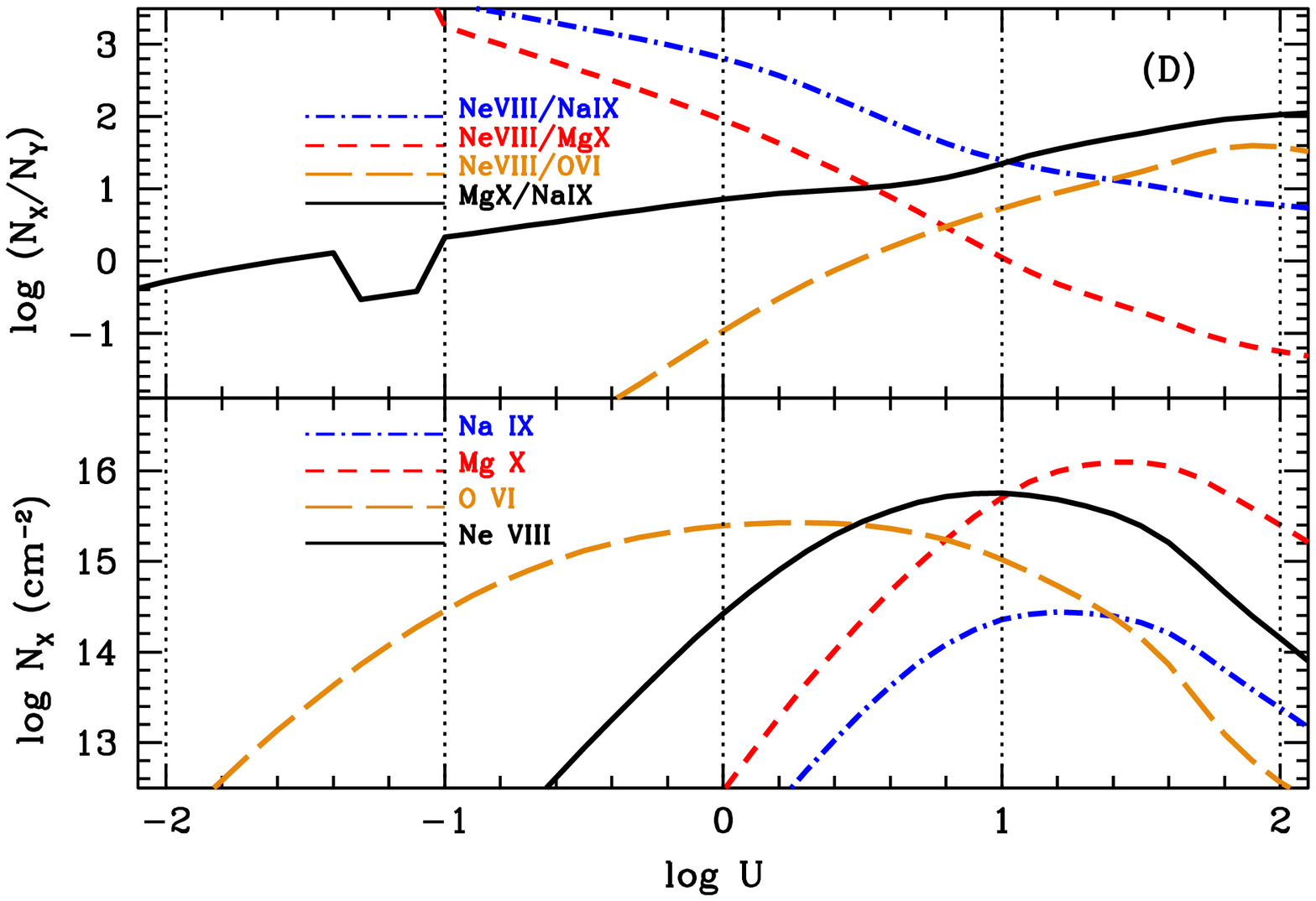} 
}}
}} 
\caption{Results of photoionization model calculation in optically thin condition with  
$N(\hi)$ = 10$^{14}$ cm$^{-2}$, incident ionizing continuum given by \citet{Mathews87}, 
and with solar metallicity $(Z = Z_{\odot})$. In each panel column densities of various 
species (bottom) and their ratios (top) are plotted as a function of ionization parameter. 
A species will be detectable for the ionization parameter range in which it has column density 
$N\gtrsim10^{13}$~cm$^{-2}$. In the detectable range if two species show constant ratio (i.e. 
insensitive to ionization parameter), they are likely to originate from the same phase of 
the absorbing gas. Such a pair of ions is good to estimate relative abundance of the elements. 
Ionization parameter should be estimated from the pair of ions whose ratio is sensitive to the 
log~U.      
Note that, according to the ionization potentials, different species are grouped and plotted in 
different panels for convenience. The notch seen in the $N(\mgx)/N(\nani)$ ratio in panel-(D) is 
an artifact created by {\sc cloudy} in the low column density limit of $N(\mgx)$.  
}   
\label{cloudy1} 
\end{figure*} 
%

In this section, we try to determine the ionization structure and the physical conditions in the outflowing gas with the help of photoionization equilibrium models using {\sc cloudy  v(07.02)} \citep[first described in][]{Ferland98}. First, we describe results of a general photoionization model to understand the variation of column densities of different high and low ions and their ratios over a wide range in ionization parameter. We then describe more detailed models (both PI and CI) only for those individual absorbers showing absorption lines from several ions with adequate column density measurements. 

Our photoionization models assume the absorbing gas to be an optically thin (i.e. stopping H~{\sc i} column density of 10$^{14}$ cm$^{-2}$ as measured in most cases) plane parallel slab with solar metallicity and relative solar abundances, illuminated by the AGN spectrum. To draw some general conclusions we use the mean spectrum of \citet{Mathews87} (hereafter MF87, see solid curve in Fig.~\ref{sed_mf}). However, it is well known that the results of photoionization modeling is very sensitive to the shape of the ionizing radiation. In order to minimize the uncertainties, while modelling individual absorbers, we use the QSO SED of the form:  
\be 
f_{\nu} = \nu^{\alpha_{\rm uv}} {\rm exp} (-h\nu/kT_{\rm BB}){\rm exp}(-kT_{\rm IR}/h\nu)+B\nu^{\alpha_{\rm x}},  
\label{agn_cont}
\en 
while discussing individual systems. Here, $T_{\rm BB}$, ${\alpha_{\rm uv}}$ and ${\alpha_{\rm x}}$ are disk black body temperature, UV spectral index and X-ray spectral index respectively. The normalization constant $B$ is fixed using the optical-to-X-ray powerlaw slope $\alpha_{\rm ox}$ and we use  $kT_{\rm IR}$ = 0.01 Rydberg. We use the SED defined by the Eq.~\ref{agn_cont} with appropriate values for the parameters (based on available observations) when we discuss the photoionization models of individual absorbers. 

The model predictions for MF87 incident continuum are plotted in Fig.~\ref{cloudy1}. In the bottom of each panel, we plot the column densities of different species having similar ionization potential, as a function of ionization parameter. For the sensitivity of our COS spectra, we find that the column density of individual species has to be $\ge10^{13}$ cm$^{-2}$ to produce detectable absorption lines which are as broad as $\sim$~100 \kms.  

In panel {\bf (A)} of Fig.~\ref{cloudy1}, we plot the model predictions for the species \nfo\ (I.P = 47.5 eV), \cf\ (I.P = 47.9 eV), \ofo\ (I.P = 54.9 eV) and \nefo\ (I.P = 63.5 eV). Among all these species, \ofo\ seems to be the dominant in the range $-2.0\le$~log~U~$\le 0.0$ and apart from \cf\ all of them showing peak round log~U~$\sim$~$-1.0$ (see bottom panel). \cf, however, shows relatively flat distribution over the above mentioned ionization parameter range. For log~U~$>0.0$, column densities of almost all these species become $<10^{13}$ cm$^{-2}$ and hence, they will not be detectable. 
From the top panel it is clear that the ionization parameter range where all the species are detectable (i.e., $-2.0\le$~log~U~$\le 0.0$), $N(\ofo)/N(\nfo)$ and $N(\ofo)/N(\nefo)$ ratios show remarkable constancy. The ratios where $N(\cf)$ is involved [i.e. $N(\nefo)/N(\cf)$ and $N(\ofo)/N(\cf)$], on the other hand, show similar constancy for $-2.0 \le$ log~U $\le -1.0$ and fall by a factor of $\ge$ 0.84 dex in the range $-1.0 \le$ log~U $\le 0.0$. 

In panel {\bf (B)} we plot the model predictions for the species \ssi\ (I.P = 72.7 eV), \of\ (I.P = 77.4 eV), \nf\ (I.P = 77.5 eV) and \nefi\ (I.P = 97.1 eV). From the bottom panel, it is apparent that \of\ is the dominant species for the whole range in ionization parameters. In addition, all of them show roughly similar $N$ distribution with a peak around log~U $\sim -0.5$. We also find that most of these species are detectable in the range $-1.5 \le$ log~U $\le 0.5$. From the top panel, it is interesting to note that, apart from $N(\of)/N(\ssi)$ ratio, all other ratios are exceptionally constant over the ionization parameter range where these species are detectable (i.e. $-1.5 \le$ log~U $\le 0.5$).                   

In panel {\bf (C)} we plot the model predictions for the species \of\ (I.P = 77.4 eV), \os\ (I.P = 113.9 eV), \arei\ (I.P = 124.3 eV) and \nesi\ (I.P = 126.2 eV). From the bottom panel, it is evident that, apart from \arei\ all other species are detectable roughly in the range $-1.0\le$ log~U $\le 1.0$. In addition, \os\ is found to be the dominant species in this ionization parameter range. \arei, on the other hand, is detectable in a very narrow range in ionization parameter around log~U $\sim 0.0$, where all these species show peak column densities. 
From the top panel, in is interesting to note that the $N(\os)/N(\of)$ ratio keeps on increasing with the increase of ionization parameter whereas $N(\os)/N(\nesi)$ ratio remains constant for the entire range in log~U (i.e. $-2.0 \le$ log~U $\le 1.0$). The $N(\nesi)/N(\arei)$ ratio also remains constant in the range $-1.0 \le$ log~U $\le 1.0$. $N(\os)/N(\arei)$ ratio, on the contrary, varies by a factor of $\gtrsim$ 6\ in the same ionization parameter range. 

In panel {\bf (D)}, we plot the model predictions for the high ionization species e.g., \os\ (I.P = 113.9 eV), \neo\ (I.P = 207.3 eV), \nani\ (I.P = 264.2 eV) and \mgx\ (I.P = 328.2 eV). From the bottom panel, we note that \nani\ and \mgx\ are detectable only for log~U~$\gtrsim$~0.5 and their column densities show peak at log~U $\sim$~1.4. \neo, on the other hand, shows peak at log~U $\sim$~1.0 and $N(\neo)>10^{13}$ cm$^{-2}$ for log~U~$\gtrsim -0.5$. \os, in contrast, shows relative flat distribution and is detectable for the entire range in ionization parameter (e.g. $-1.5 \le$ log~U $\le 1.8$). 
It is interesting to note that the ratios plotted in the top panel show smooth variation over the whole range in ionization parameter. For example, $N(\mgx)/N(\nani)$ ratio varies by a factor $\sim$ 3\ in the range $0.0 \le$ log~U $\le 1.0$ and by a factor $\sim$ 5\ in the range 1.0 $\le$ log~U $\le$ 2.0. Note that notch seen in $N(\mgx)/N(\nani)$ ratio around log~U = $-1.2$ is not real but a numerical artifact  where $N(\mgx)$ become 
negligibly small. 

The above analysis clearly provides the rough range in the ionization parameter where species with similar ionization potentials are most likely to originate from the same phase of the absorber. In this U range the ratios of such ionic column densities are also useful in constraining the relative abundances of the heavy elements. On a different note, we wish to point out here that all these species originating from same phase (or density) will have similar projected area and hence they will show very similar covering fractions.    
The ratios of very highly ionized species (i.e. \os, \neo, \nani\ and \mgx\ that are the main focus of this work) show smooth variation over ionization parameter. These ratios are sensitive probes of the ionization parameter provided these
species originate from the same phase of the absorbing gas. The nature of absorption profiles (e.g. velocity alignment, line spread, component structure etc.) can  be used to decide whether these species originate from the same phase of the absorbing gas.    

\begin{figure} 
\centerline{
\vbox{
\centerline{\hbox{ 
\includegraphics[height=8.4cm,width=8.4cm,angle=00]{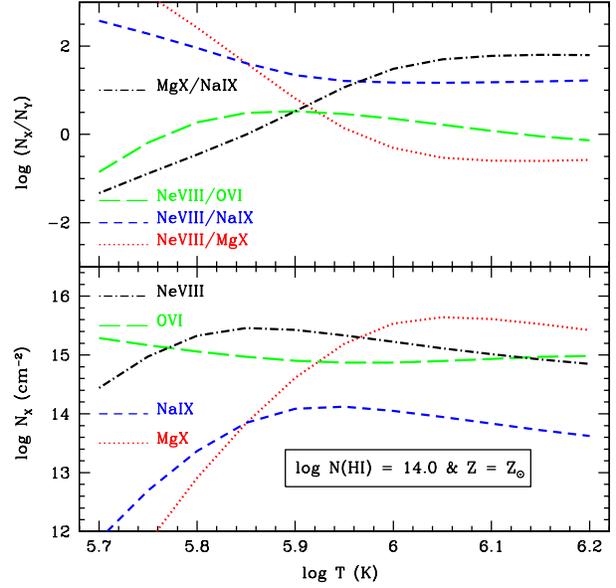} 
}}
}}
\caption{ {\sl Bottom :} Column densities of various high ionization species 
as a function of gas temperature under collisional ionization equilibrium 
\citep[]{Sutherland93}. Column densities are calculated for $N(\hi) = 10^{14}$ 
cm$^{-2}$, assuming solar metallicity.     
{\sl Top :} Column density ratios are plotted as a function of gas temperature.}    
\label{CIE_ratio} 
\end{figure} 
%
 
\citet{Muzahid12b} have shown that the near constancy of $N$(O~{\sc vi})/$N$(Ne~{\sc viii}) between different components in the associated absorber towards HE~0238--1904 can be explained if collisional excitation plays an important role. Therefore, we  now consider the collisional ionization equilibrium (CIE) model \citep[]{Sutherland93}. The model predicted column densities of high ionization species discussed in the panel-{\bf (D)} of  Fig.~\ref{cloudy1} are plotted as a function of gas temperature, in the lower panel of Fig.~\ref{CIE_ratio}. The column densities are calculated for $N(\hi)$ = 10$^{14}$ cm$^{-2}$ and $Z = Z_{\odot}$, typically seen in most of the cases in our sample. It is clear from the figure that for log~$T >$ 6.0, all these high ionization species become fairly insensitive to the gas temperature. This fact is also manifested in the column density ratios, plotted in the top panel.  

In what follows we provide detailed models for some individual systems (specially the ones that show \nani) in the framework of photoionization and CIE models.

\subsection{Models for the system \zabs\ = 1.02854 towards PG~1206$+$459}
\label{sec_phot_model_pg1206} 

%
\begin{figure} 
\centerline{
\vbox{
\centerline{\hbox{ 
\includegraphics[height=8.4cm,width=8.8cm,angle=00]{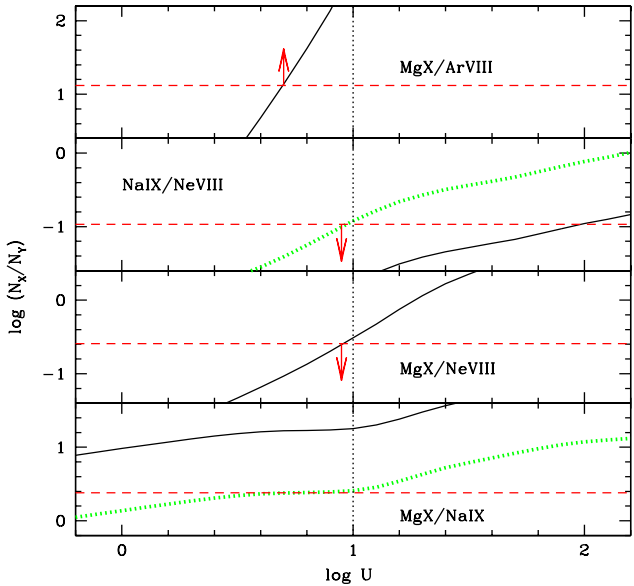} 
}}
}}
\caption{Photoionization model for the system \zabs\ = 1.02854 towards PG~1206$+$459.  
{\sc cloudy} predicted column density ratios of various high ionization species as 
a function of ionization parameter are plotted in different panels. The horizontal 
dashed line in the bottom  panel indicates the measured value of $N(\mgx)/N(\nani)$ 
in component-1. In all other cases the horizontal dashed line marks the upper/lower 
limit on the ratio as shown by an arrow.    
The dotted (green) curves are the model prediction in case of $\rm Na$ is overabundant by 
a factor of 7 relative to $\rm Mg$ and/or $\rm Ne$. The dotted vertical line represents a 
possible solution for the ionization parameter. 
} 
\label{model_pg1206} 
\end{figure} 
%

\begin{figure*} 
\centerline{
\vbox{
\centerline{\hbox{ 
\includegraphics[height=9.2cm,width=9.2cm,angle=00]{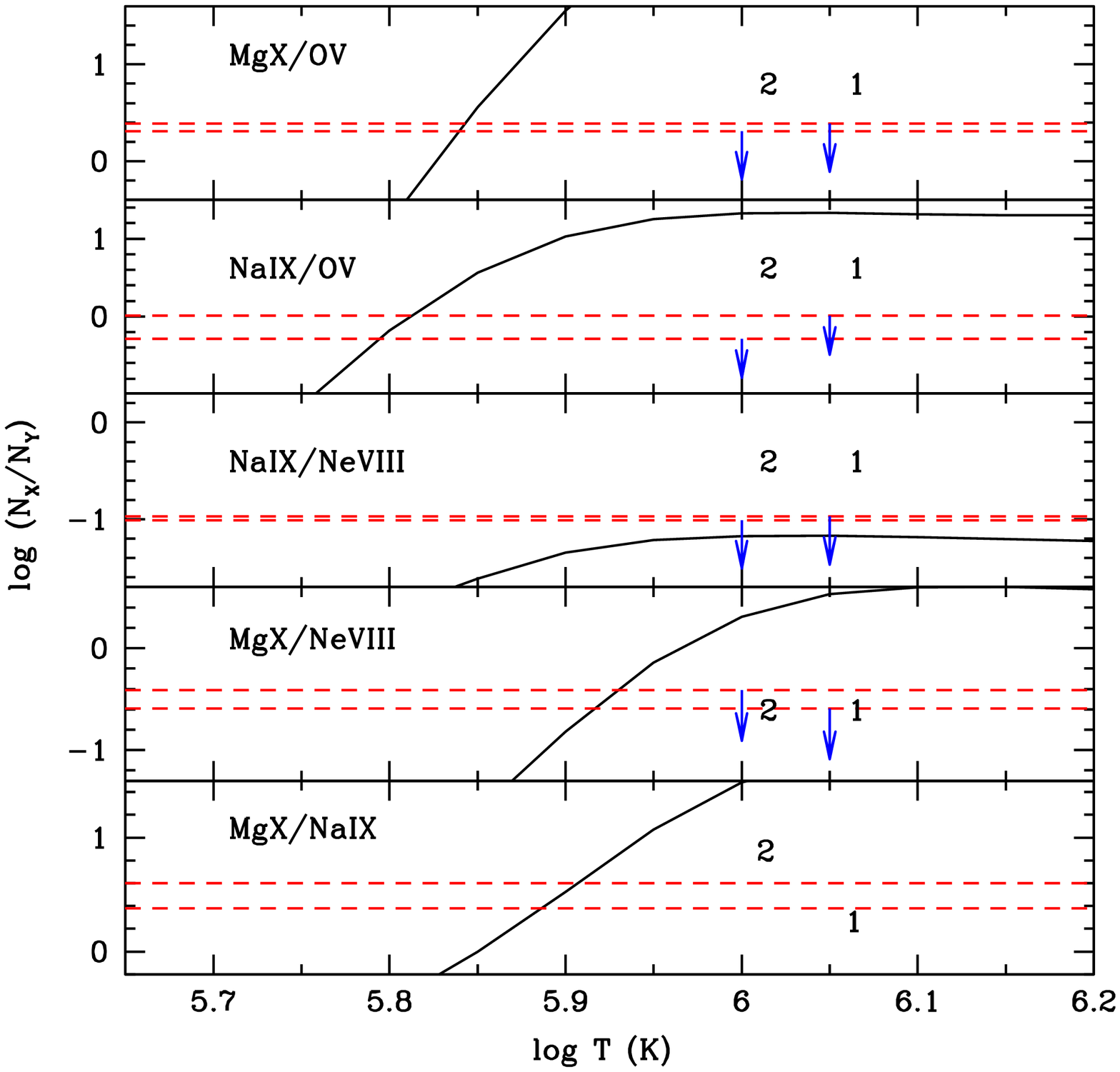} 
\includegraphics[height=9.2cm,width=9.2cm,angle=00]{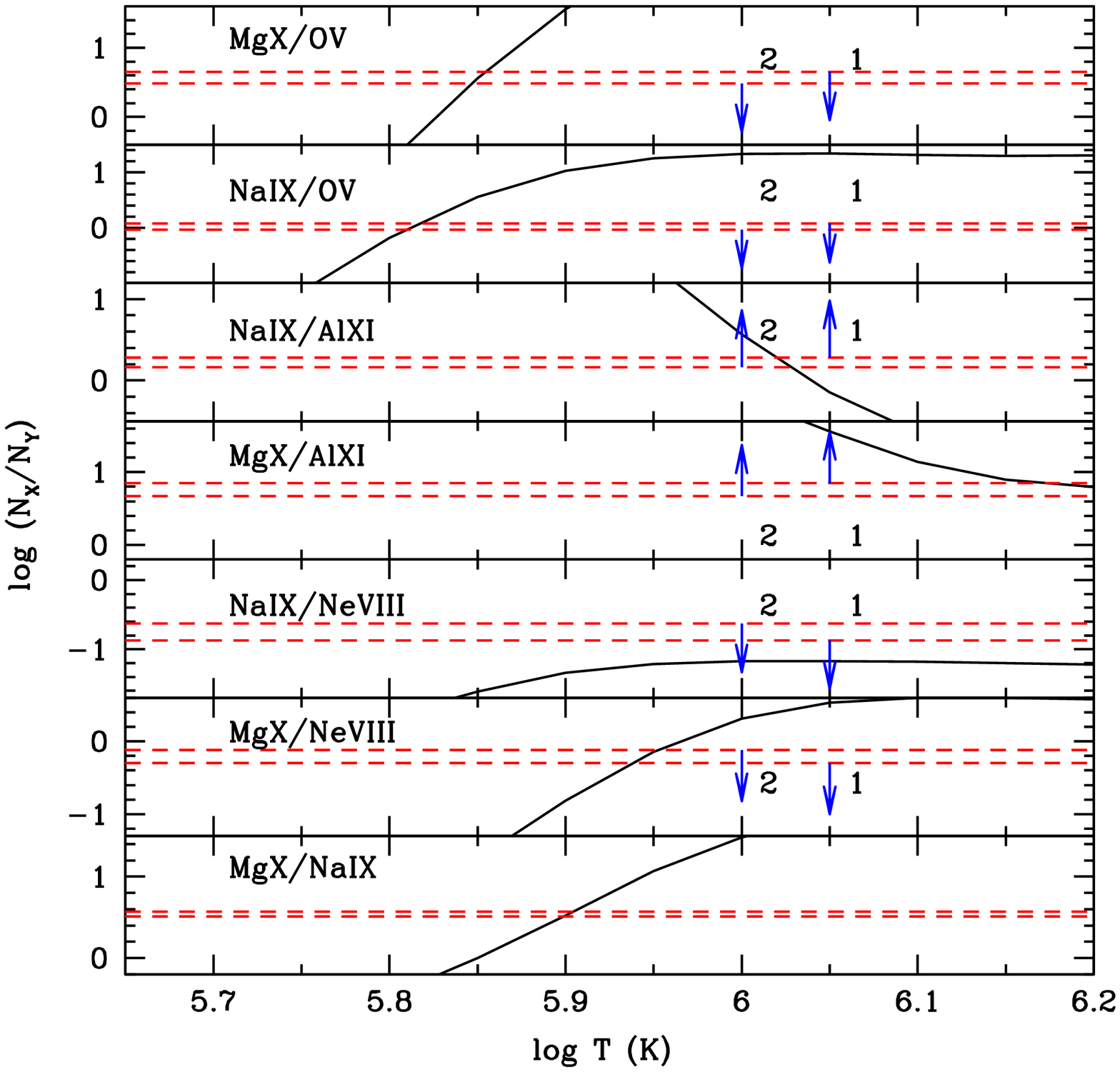} 
}}
}}
\caption{{\sl Left:} CIE model for the system at \zabs\ = 1.02854 towards PG~1206$+$459. 
{\sl Right:} CIE model for the system at \zabs\ = 1.15456 towards PG~1338$+$416. 
In each panel model predicted column density ratios of various high ionization species 
are plotted as a function of gas temperature. In the bottommost panels the horizontal dotted 
line represents the measured $N(\mgx)/N(\nani)$ ratios in component-1 and component-2. In all 
other panels horizontal dotted lines followed by an arrow  mark the observed upper/lower limits 
on the plotted ratios.
} 
\label{CIE_model} 
\end{figure*} 

Here we discuss  the physical conditions in the \zabs\ = 1.02854 system towards PG~1206$+$459. The ionizing background is characterized by the Eq.~\ref{agn_cont} with  $T_{\rm BB} \sim 1.5 \times 10^{5}$~K, $\alpha_{ox} = -1.7$, $\alpha_{x} = -0.7$ and   $\alpha_{uv} = -0.5$. 
The value of $\alpha_{ox}$ has been calculated assuming power law shape of X-ray spectrum with photon index, $\Gamma_{\rm 0.3-12 keV} = -1.74\pm0.09$ (or $\alpha_{x} \sim -0.7$) and normalization at 1keV is $A_{\rm PL} = (2.4\pm0.2)\times10^{-4}$ $\rm photons~cm^{-2} keV^{-1} s^{-1}$, as estimated for this source by \citet{Piconcelli05}. Using the black hole mass, M$_{\rm BH} = 1.0 \times10^{9}$ M$_{\odot}$, and $\rm L_{Bol}/L_{Edd} = 0.84$ from \citet{Chand10}, the inner disk temperature ($T_{\rm BB}$) is found to be very similar to the value used here. 

From Fig.~\ref{vp_pg1206} (and Table~\ref{tab_pg1206}) we notice that the covering fractions for \nani, \neo, \of, \mgx\ and \arei\ are similar. It is clear that \neo\ and \of\ column density estimation are lower limits as they are affected by saturation effects. The photoionization model predictions for the above mentioned SED is given in Fig.~\ref{model_pg1206}. The horizontal dashed line in each panel represents the observed values for the component 1 (i.e., $v_{\rm ej} \sim -19,250$ \kms component in Table~\ref{tab_pg1206}).  The upper limit on the observed $N(\mgx)/(\neo)$ ratio, suggests log~U~$\le$~1.0. The lower limit on the observed $N(\mgx)/(\arei)$ ratio, on the other hand suggests log~U~$\ge$~0.7. We notice that in this ionization parameter range (i.e. 0.7 $\le$ log~U $\le$ 1.0) the model over-predicts the observed $N(\mgx)/N(\nani)$ ratio. The observed column density ratios involving \nani\ can be reproduced by the models if we assume ${\rm Na}$ is enhanced by a factor of 0.85 dex with respect to ${\rm Mg}$ and/or ${\rm Ne}$ [see the dotted (green) curves in Fig.~\ref{model_pg1206}].
We estimate the upper limit for  log~$N(\hi)$ (cm$^{-2}$) = 13.78 using $f_c = 0.59$ (see Table~\ref{tab_pg1206}). For this model predicts log~$N$(\mgx) (cm$^{-2}$) = 15.32, which is very close to the observed value implying the metallicity of the gas phase producing \neo\ and \mgx\ is higher than solar.
Among the other species detected in this component only \ofo\ column density and covering fraction are well measured. We find $N$(\ofo) predicted by our model for log~U$\sim 1$ is a factor 25 times smaller than what is observed. This confirms that \ofo\ is originating from a distinctly different phase as suggested by the low covering fraction as well. The observed column density of \ofo\ for solar metallicity and log~$N(\hi)$ (cm$^{-2}$) = 14.42 (for the similar covering fraction measured for \ofo) we find log~U $\sim 0$. This ionization parameter also produces the correct value of observed $N(\nfo)$. If both these phases are at the same distance from the QSO then we can conclude that there is a factor ten change in the density along the transverse direction for the absorbing gas. 

In the case of component-2 (i.e., $v_{\rm ej}\sim -19,150$ \kms\ component in Table~\ref{tab_pg1206}), the upper limit on observed $N(\mgx)/(\neo)$ ratio, suggests log~U~$\le$~1.0. The lower limit on observed $N(\mgx)/(\arei)$ ratio,  on the other hand suggests log~U~$\ge$~0.8. As in the case of component-1 for this ionization parameter range the photoionization model over predicts the observed $N(\mgx)/N(\nani)$.  We find that the observed column density ratios involving \nani\ can be reproduced by the model if we assume $\rm Na$ is enhanced by a factor of 0.60 dex with respect to $\rm Mg$ and/or $\rm Ne$. Like in the previous case the model that reproduces the high ions under-predicts the \ofo\ column density. We also find \ofo\ is originating from a phase that is up to a factor 10 lower density if both phases are at same distance from the QSO. 

In both the components \nf\ absorption is detected. The measured column densities are consistent with an ionization parameter intermediate between the gas traced by \nfo/\ofo and \mgx. All this suggests that the outflow having smooth density gradients in the transverse direction.

For log~U=1, the inferred total column density of system is $N(\rm H)$ = 4.7~$\times$~10$^{20}$ cm$^{-2}$ (when log~$N(\hi)$ (cm$^{-2}$) = 14.0), whereas, \ose\ and \oei\ column densities are $N(\ose)$~ = 1.0~$\times$~10$^{17}$ cm$^{-2}$ and $N(\oei)$~ = 9.5~$\times$~10$^{16} \rm cm^{-2}$, suggesting continuum optical depths of \ose\ and \oei\ are much less than 0.1. Therefore, this system may not be a potential X-ray WA candidate. 

In the left hand panel of Fig.~\ref{CIE_model}, the column density ratios of  various high ionization species predicted by the CIE models, are plotted as a function of gas temperature. The horizontal dashed lines followed by arrows, in each sub-panel except for the bottom one, indicate the upper limit on the column density ratios measured in component-1 and component-2. The measured values of $N(\mgx)/N(\nani)$ ratio, shown in the bottom panel, are found to be very similar for both the components which corresponds to a temperature of log~$T \sim 5.9$. Note that the upper limits on $N(\mgx)/N(\neo)$ ratios observed in both the components suggests log~$T \lesssim 5.9$. The observed value of $N(\of)$, on the other hand, suggests a temperature $T\sim10^{5.8}$~K. On the other hand we notice that low ionization species like O~{\sc iv} and N~{\sc iv} require $T\sim10^{5.2}$ K. In order for the two phases to be in pressure equilibrium the density of the low ionization phase needs to be a factor 4 higher. We next run {\sc cloudy} model keeping the gas temperature to be constant at $T\sim10^{5.8}$~K and found that the ionization parameter of the gas log~U$\le-2$ so that the ratio of \mgx\ and \neo\ are not affected by the QSO radiation. Given the luminosity of the QSO this corresponds to a radial separation of $\gtrsim 2600/\sqrt{(n_{\rm H}/10^5)}$~pc between the absorbing gas and the QSO, so that the ionization state can be dominated by collisions.  

\subsection{Models for the system \zabs\ = 1.15456 towards PG~1338$+$416} 
\label{sec_phot_model1_pg1338}

\begin{figure} 
\centerline{
\vbox{
\centerline{\hbox{ 
\includegraphics[height=9.2cm,width=9.2cm,angle=00]{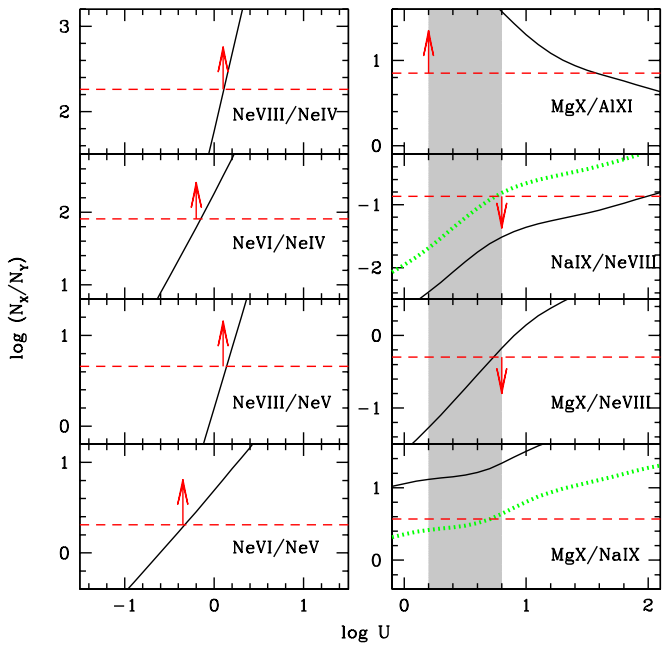} 
}}
}}
\caption{Photoionization model for the system \zabs\ = 1.15456 towards PG~1338$+$416. 
{\sc cloudy} predicted column density ratios of different ions are plotted as a 
function of ionization parameter in different panels. The horizontal (red) dashed line 
followed by an arrow in each panel represents the observed upper/lower limit on the 
plotted ratio as measured in component-1. 
{\sl Left :} Ratios of different ionization states of neon are plotted. 
{\sl Right :} Same as {\sl Left} but for different ionization states of different elements. 
The dotted (green) curves are the model prediction in case of $\rm Na$ is overabundant by a 
factor of 5 relative to $\rm Mg$ and/or $\rm Ne$. The shaded region represents the allowed 
range in ionization parameter as suggested by various ionic ratios.             
} 
\label{phot_pg1338} 
\end{figure} 

Here we discuss the photoionization model for the system \zabs\ = 1.15456 towards PG~1338$+$416. For SED we use $T_{\rm BB}\sim1.0\times10^{5}$~K, $\alpha_{\rm x} = -1.5$, $\alpha_{\rm uv} = -0.5$ and $\alpha_{\rm ox} = -1.8$ \citep[from][]{Anderson07}. From the \mgt\ emission line width \citet{Chand10} have estimated the black hole mass for this source to be log~M$_{\rm BH}$/M$_{\odot} \sim$ 8.96 and 9.47 using the method by \citet{McLure04} and \citet{Dietrich09} respectively. They also find $L_{\rm bol}/L_{\rm Edd} = 0.34$ for this source. Using these we calculate the inner disk temperature for this QSO to be $T_{\rm BB} \sim 1.2 \times 10^{5}$~K and $9.1 \times 10^{4}$~K for log~M$_{\rm BH}$/M$_{\odot} \sim$ 8.96 and 9.47 respectively. This is close to what we use to generate the SED.  

In Fig.~\ref{phot_pg1338} we show  the results of our photoionization model. In the left hand panel of the figure we have shown the column density ratios of different ionization states of neon. All these ratios gives lower limits on ionization parameter. The best constraint comes from $N(\neo)/N(\nefi)$ ratio, which suggests log~U~$\ge$~0.2. All other ratios are consistent with this lower limit. In the right hand panel we have plotted ionic ratios of different species of different elements which can provide useful constraints on the ionization parameter (see section \ref{sec_phot_model}).  The observed upper limit on $N(\mgx)/N(\neo)$ ratio suggests log~U~$\le$~0.8. Hence the physically allowed range in ionization parameter becomes $0.2 \le$~log~U~$\le 0.8$, as marked by the shaded region. We note that the observed limits on $N(\mgx)/N(\alel)$ and/or $N(\mgx)/N(\of)$ are also consistent with this range. However, it is apparent from the right-bottom panel, that our model cannot reproduce the observed $N(\mgx)/N(\nani)$ ratio for the whole range in ionization parameter, where the individual species (i.e. \nani\ and \mgx) are detectable. Similarly, $N(\nani)/N(\neo)$ ratio also suggests a very high log~U, which is not in the allowed range of ionization parameter (i.e. shaded region). Like the previous case (see section~\ref{sec_phot_model_pg1206}), such a  discrepancy can be easily avoided if $\rm Na$ is overabundant by factor of $\sim$~5--6, as can be seen from the dotted (green) curves in the figure. 

Assuming log~U = 0.5 and using the estimated upper limit on $N(\hi)$ in the component-1 [i.e. log~$N(\hi)$ (cm$^{-2}$) $\le$ 13.64; see Table~\ref{tab1_pg1338}], we estimate the metallicity of the gas to be $\gtrsim$ 10 $Z_{\odot}$. The total column density of the system at log~U~$\sim$~0.5 is log~$N(\rm H)$ (cm$^{-2}$) = 20.09. Predicted column densities of \ose\ and \oei\ are log $N$ (cm$^{-2}$) = 17.47 and 17.02 respectively. Continuum optical depth of oxygen corresponding to these values is again much less than 0.1, suggesting that the system may not be a potential X-ray WA candidate.   

From the right hand panel of Fig.~\ref{CIE_model}, we can conclude that the observed ratios and limits of high ions can be explained if the gas temperature is $T\sim 10^{5.9}$ K without the enhancement of $\rm Na$ as required by the photoionization models. However, in order for the QSO radiation field to not affect the ionization state of the absorbing gas the ionization parameter has to be log~U~$\le-1.0$. For the inferred luminosity this corresponds to a separation of $\gtrsim400/\sqrt{(n_{\rm H}/10^5)}$~pc of the absorbing cloud from the QSO.  

\subsection{Model for the system \zabs\ = 1.21534 towards PG~1338$+$416} 
\label{sec_phot_model2_pg1338} 

In this section we discuss the photoionization model for \zabs\ = 1.21534 absorber towards PG~1338$+$416. This system has \zabs\ very similar to \zem\ = 2.2145$\pm$0.0019. This is the only associated \neo\ system along the line of sight without detectable \nani\ absorption. Unlike this system the other two have large outflow velocities and show signatures of partial coverage. The column density of \neo\ is also high in the other two systems.

\begin{figure} 
\centerline{
\vbox{
\centerline{\hbox{ 
\includegraphics[height=7.2cm,width=8.8cm,angle=00]{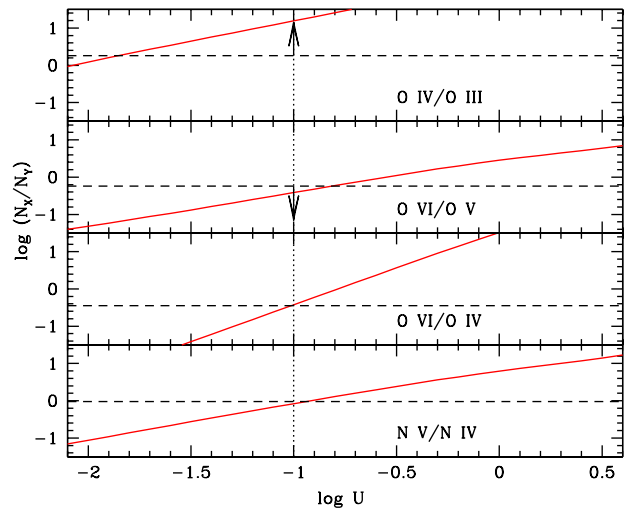}  
}}
}}
\caption{Photoionization model for the system \zabs\ = 1.21534 towards PG~1338$+$416. 
{\sc cloudy} predicted column density ratios of different ions are plotted as a function 
of ionization parameter in different panels. The horizontal dashed line in the bottom two 
panels mark the measured ionic ratios. The dashed lines with arrows in the top two panels 
show measured limits on the ionic ratios. The vertical dotted line at log~U = $-1$ marks 
a possible solution for the ionization parameter.   
} 
\label{fig:phot_1.21534_pg1338} 
\end{figure}

In section~\ref{sec_discript_PG1338_1.21534}, we have seen that the low ionization species, detected in COS, originating from this system show 2 possible components. However, because of the poor spectral resolution, species detected in $HST/$FOS spectra can be well fitted by a single Voigt profile component. Because of this disparity in the data quality, we use the total column densities (i.e. summed up component column densities), for the photoionization model. We run {\sc cloudy} with same set of parameters as described in section~\ref{sec_phot_model1_pg1338}. The results of our photoionization model are shown in Fig.~\ref{fig:phot_1.21534_pg1338}. 

The column density ratios of different ionization states of same element are very important diagnostics of ionization parameter. Therefore, we make use of simultaneous presence of \oth, \ofo, \of\ and \os\ lines of oxygen and \nfo\ and \nf\ lines of nitrogen to estimate the ionization parameter of the gas. Since we treat $N(\oth)$ and $N(\of)$ as upper and lower limits (see discussions in section~\ref{sec_discript_PG1338_1.21534}), the ionization parameter is primarily decided by  $N(\os)/N(\ofo)$ and $N(\nf)/N(\nfo)$ ratios. It is clear from Fig.~\ref{fig:phot_1.21534_pg1338} that, both the ratios are remarkably consistent with log~U $\sim -1.0$. We also note that, the upper limit on $N(\os)/N(\of)$ and lower limits on $N(\ofo)/N(\oth)$ ratios are also suggestive of such an ionization parameter. Using the ionization fractions at log~U $\sim -1.0$, we find that the metallicity of the gas to be near solar, e.g. log~$Z/Z_{\odot}$ = 0.40$^{+0.90}_{-0.25}$. 

In the section~\ref{sec_phot_model}, we have seen that the species \nfo\ and \ofo\ traces each other for a wide range in ionization parameter. Therefore, $N(\nfo)/N(\ofo)$ ratio is a sensitive probe of the relative abundances. From the observed $N(\nfo)/N(\ofo)$ ratio we find that nitrogen is overabundant compared to oxygen by a factor of 0.93 dex (i.e., $\rm [N/O] = 0.07$). Furthermore, nitrogen is found to be overabundant compared to carbon by a factor of 0.89 dex (i.e. $\rm [N/C] = 0.29$), from the measured  $N(\nfo)/N(\ct)$ ratio. Since \oth\ line is blended, we use $N(\ct)/N(\ofo)$ ratio to estimate $\rm [C/O]$ and found that the carbon and oxygen roughly follow solar abundance pattern. For example, estimated $\rm [C/O] = -0.22$, whereas, in sun $\rm (C/O) = -0.26$ \citep[]{Asplund09}. Such an enhanced nitrogen abundance is seen in high redshift ($z\ge2.0$) QSOs \citep[]{Hamann92,Korista96,Petitjean99}. These authors suggested a rapid star formation scenario which produces a super solar metallicity in order to boost the nitrogen abundance through enhanced secondary production in massive stars.  We would like to mention that, with the estimated ionization parameter and metallicity, neither \neo\ nor \mgx\ would be detectable [e.g., reproduced log~$N(\neo)$ (cm$^{-2}$) $\ll$ 14.0 at log~U = $-1.0$].     

The observed $N(\neo)/N(\mgx)$ ratio require a different phase with fairly high ionization parameter (i.e., log~U~$\sim$~1.3). If we assume  most of the \of\ originate from \neo\ phase then we get log~U $\ge$~0.8. In equality in this case is because some part of  \of\ will originate from \ofo\ phase. If we use $N$(\os)/$N$(\neo) ratio then we get log~U $\ge$~0.9. Thus one can conclude that the \neo\ absorption is originating from a gas having log~U~$\sim$ 1 (as we have seen in the other cases discussed above).
If we assume the \neo\ phase has same metallicity as the low ionization phase discussed above then we can conclude that \hi\ associated with \neo\ is $\le 10^{12}$ cm$^{-2}$. We can conclude that the low hydrogen column density in this component  is the reason for the lack of \nani\ absorption in this system. 

Like in the previous cases our model suggests that the absorbing gas will not have sufficient optical depth to be a X-ray warm absorber. 

\begin{table*}
\caption{Summary of associated \neo\ absorbers from literature (only secure detections are listed here)}  
\centering  
\begin{tabular}{crrcccccccc} 
\hline 
    &      &        &     \multicolumn {5}{c} {log~$N$~(cm$^{-2}$)} & \\ \cline{4-8}   
QSO & \zem & $v_{\rm ej}$(\kms) &  \os  & \neo & \mgx &  \hi & $\rm H$ & Type &  QSO Type &  Reference \\ \hline   
 (1)  &     (2)    &     (3)     &    (4)      &    (5)    &  (6)    &   (7)   &    (8)       &    (9)   & (10)  & (11)  \\   
\hline 
UM~675         & 2.150 & $-$1500  & 15.5  &  15.4  &  ....  &  14.8  &  20.0  &  NAL  &  RQ$^{b}$  &  \citet{Hamann95} \\ 
SBS~1542$+$541 & 2.631 & $-$11360 & 15.8  &  16.0  &  15.9  &  14.9  &  22.7  &  BAL  &  RQ        &  \citet{Telfer98} \\ 
J~2233$-$606   & 2.240 & $-$3900  & 15.4  &  15.1  &  ....  &  14.0  &  22.0  &  NAL  &  ...       &  \citet{Petitjean99} \\ 
PG~0946$+$301  & 1.221 & $-$10000 & 16.6  &  16.7  &  16.6  &  15.3  &  ....  &  BAL  &  RQ        &  \citet{Arav99a} \\ 
3C~288.1       & 0.965 & $+$250   & 15.8  &  15.4  &  15.0  &  15.8  &  20.2  &  NAL  &  RL$^{a}$  &  \citet{Hamann00} \\ 
\hline 
\hline 
\end{tabular}
~\\ 
Table Note -- $^{a}$ Radio Loud; ~~$^{b}$ Radio Quiet 
\label{tab:summary} 
\end{table*}  
%

%
\begin{figure} 
\centerline{\hbox{ 
\centerline{\vbox{
\includegraphics[height=8.0cm,width=8.0cm,angle=00]{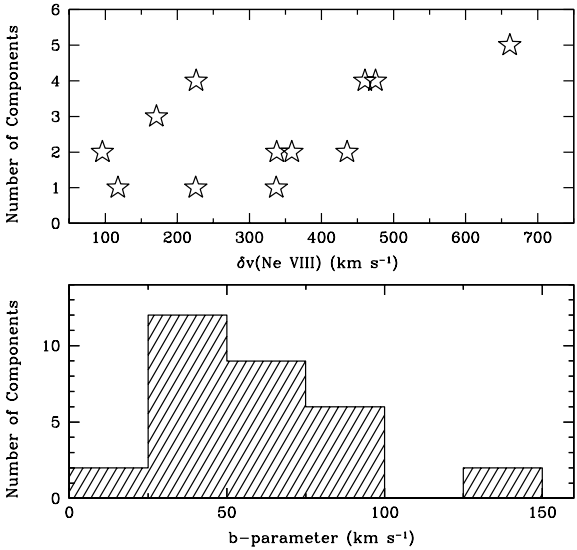} 
}}
}}
\caption{ {\sl Bottom:} The distribution of Doppler parameter as measured in individual 
\neo\ components. {\sl Top:} Number of \neo\ components against the spread of \neo\ 
absorption in each system.      
} 
\label{b_dist} 
\end{figure} 
%
 
\section{Discussions} 
\label{sec_diss}

In this section, we try to draw a broad physical picture of the associated \neo\ absorbers.

\subsection{Incidence of associated \neo\ absorbers} 

The fraction of AGNs that show associated absorption is important for understanding the global covering fraction and the overall geometry of the absorbing gas \citep[]{Crenshaw03,Ganguly08}. In low redshift Seyfert galaxies, surveys in the UV \citep[]{Crenshaw99}, FUV \citep[]{Kriss02}, and X-rays \citep[]{Reynolds97a} found $\sim$ 50 -- 70\% incidence of associated absorbers. For quasars, the fraction of occurrence has been found to be somewhat lower. For example \citet[]{Ganguly01} found signature of associated \cf\ absorption in $\sim$~25\% of QSOs. On the other hand, \citep[]{Dai08} have found occurrence of BAL in $\sim$~40\% cases. However, we note that depending upon the selection criteria (e.g. cutoff velocity, rest frame equivalent width etc.) these numbers could be very different. An exposition on the incidence of different forms of the associated absorbers can be found in \citet{Ganguly08}. We have found 12  associated \neo\ systems in 8 out of 20 QSOs in our sample while only 2 is expected based on the statistics of intervening systems. Even if we restrict ourself to $|v_{\rm ej}|$ up to 5000 \kms\ instead of 8000 \kms, we have 8 associated systems which is factor 4 higher compared to what is expected from statistics of intervening systems. Such an enhanced occurrence of associated absorbers have also been noticed in the case of high-$z$ \citep[]{Fox08} and low-$z$ \citep[]{Tripp08} \os\ absorbers.     
The incidence of associated \neo\ absorbers in our sample is $\sim$40\% ($\sim$35\% if we do not include the tentative system towards HB89 0107$-$025 or restrict to systems with $|v_{\rm ej}|<$ 5000 \kms). It is also interesting to note only 5/12 systems along 3/20 sightlines show signature of partial coverage. Therefore the incidence of partially covered associated \neo\ absorber is 15\%. No associated \neo\ system is detected towards 7 radio bright QSOs in our sample. There are 5 \neo\ absorption reported in the literature (see Table~\ref{tab:summary}) and only one of them (\zabs = 0.965 towards 3C~288.1) is towards radio bright QSO. Confirming the high detection rate of associated \neo\ systems and relatively less incidence rate towards radio bright QSOs is very important to understand the possible influences of radio jets.

\subsection{Line broadening}  
In the bottom panel of Fig.~\ref{b_dist} we show the distribution of Doppler parameter as measured in individual \neo\ components. The median value of $b(\neo)$ is $\sim$~58.7 \kms. The upper limit on temperature corresponding to this value is 10$^{6.6}$~K. Under CIE, even \neo\ will not be a dominant species at such high temperatures. The collisional ionization fraction of \neo\ becomes only $\sim 3\times10^{-3}$ at $T \sim 10^{6.6}$~K. Therefore the width of individual Voigt profile components are most probably dominated by non-thermal motions. We note that $b(\neo) \sim$~22~\kms\ corresponds to a temperature of 10$^{5.8}$~K, at which $N(\neo)$ peaks under CIE (see bottom panel of Fig.~\ref{CIE_ratio}). This indeed suggests that, based on the observed $b$-values of \neo, we cannot rule out the possibility of gas temperature being 6--7$\times$~10$^{5}$~K at which collisional ionization becomes important.  
Note that we use minimum number of Voigt profile components needed to have a reduced $\chi^2\sim 1$. The discussions presented above are based on $b$-parameters derived this way. While we can not rule out each of our Voigt profile component being made of a blended large number of components, our analysis suggests that the observed line profiles allow for the gas temperature being higher than the typical photoionization equilibrium temperature.     
In the top panel of Fig.~\ref{b_dist} we have plotted the number of components required to fit \neo\ absorption against the velocity spread of the line. Lack of any significant correlation between these two suggests that the line spread may not dominated by the presence of multiple number of narrow components \citep[as seen in the case of high redshift \os\ absorbers, e.g.][]{Muzahid12a}  but the line spread is related to the large scale velocity field. Further, $\delta v (\neo)$ lies roughly between 100 -- 800~\kms\ suggesting that these absorbers are intermediate of BAL and NAL. This type of associated absorbers are also known as mini-BAL. 

\subsection{Ejection velocities and correlations} 
\begin{figure} 
\centerline{\hbox{ 
\centerline{\vbox{
\includegraphics[height=10.0cm,width=8.6cm,angle=00]{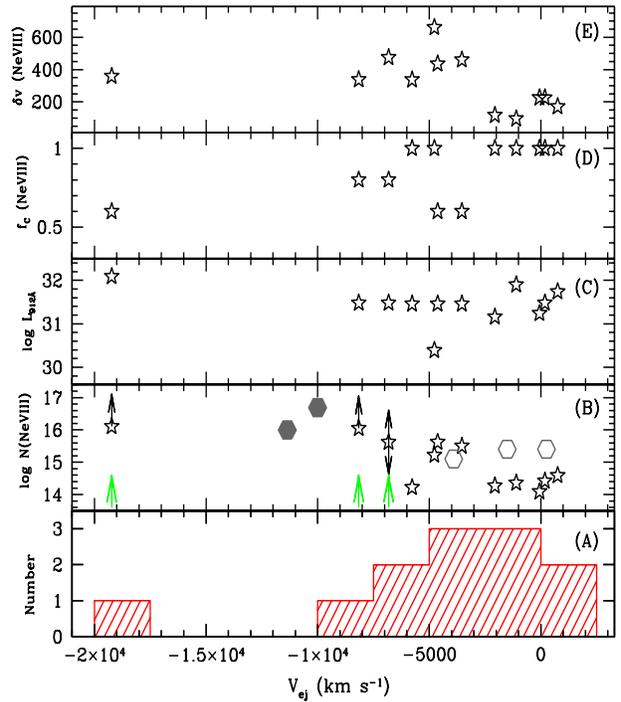} 
}}
}}
\caption{{\bf (A):} The distribution of ejection velocity for the associated \neo\ 
absorbers presented in this paper. {\bf (B):} The \neo\ column density as a function 
of ejection velocity. The hexagons are from literature listed in Table~\ref{tab:summary}. 
The filled hexagons represent BALQSOs. The (green) upward arrows mark the systems where 
we detect \nani. {\bf (C):} $L_{912\AA}$ as a function of $v_{\rm ej}$. {\bf (D):} \neo\ 
covering fractions in individual systems against $v_{\rm ej}$. {\bf (E):} The line spread 
of \neo\ absorbers against $v_{\rm ej}$.    
} 
\label{beta_dist} 
\end{figure}

The ejection velocity is defined as the velocity separation between the emission redshift of the QSO and the \neo\ optical depth weighted redshift of the absorber. The distribution of ejection velocities in our sample are shown in panel {\bf (A)} of Fig.~\ref{beta_dist}. Clearly most of these associated \neo\ absorbers are detected within $-5000$ \kms\ from the emission redshift of the QSO.  The highest velocity absorber is detected at a ejection velocity of $\sim-19,000$ \kms. In panel {\bf (B)} we have plotted covering fraction corrected total column densities of \neo\ in our sample ({\sl stars}) as a function of ejection velocity. The hexagons in this panel are from literature (see Table~\ref{tab:summary}). The overall sample shows a possible correlation between $N(\neo)$ and $v_{\rm ej}$. If we consider all the limits as detections we find a 2.1$\sigma$ correlation for the systems in our sample. When we consider the measurements from the literature the significance of the correlation increase to 2.7$\sigma$. However, we note that the top two ejection velocity systems from the literature (filled hexagons) are BAL in nature. The (green) arrows in the bottom, identify the systems with \nani\ detection. It is apparent that these are the ones having top three ejection velocities with $|v_{\rm ej}| > 5000$ \kms, in our sample. Interestingly, we note that the only possible \nani\ detection was reported before, by \citet{Arav99a} towards BALQSO PG~0946$+$301 where the system has an ejection velocity of $-10,000$ \kms, which is consistent with the trend seen in our sample.   

In panel {\bf (C)} we have plotted the Lyman continuum luminosity (i.e. $L_{912\rm \AA}$ in ergs s$^{-1}$ Hz$^{-1}$) of the sources in our sample as a function of $v_{\rm ej}$. We do not find any obvious correlation between them. However, the highest velocity system, which also show \nani\ absorption, originates from the highest UV luminosity source. Here we note that, the sources with higher UV luminosities are found to be the ones with higher outflow velocities in the sample of SDSS BALQSOs \citep[see e.g.][]{Gibson09}. 
The estimated \neo\ covering fractions in different systems in our sample are plotted against the ejection velocity in panel {\bf (D)}. It is to be noted that majority of the systems at smaller ejection velocities show nearly 100\% coverage of the background source whereas the systems with higher ejection velocity tend to have lower covering fractions. In panel {\bf (E)} the line spreads of \neo\ absorption in each system are plotted as function of $v_{\rm ej}$. A mild 2$\sigma$ level correlation is seen between $\delta v(\neo)$ and $v_{\rm ej}$, suggesting systems with higher outflow velocity are likely to show wider spread. 

%
%

%
\begin{table}
\vfill 
\caption{List of intervening \neo\ absorbers that exist in literature}  
\begin{tabular}{cccccc} 
\hline 
      &       &       & \multicolumn{2}{c}{log~$N$ (cm$^{-2}$)} & \\ \cline{4-5}
  QSO & \zem\ & \zabs &       \neo    &       \os      &  Ref.$^{a}$   \\  \hline  
  (1) &  (2)  &  (3)  &       (4)     &       (5)      &    (6)        \\ 
\hline 
\hline 
PG~1148$+$549    &  0.9754  &  0.6838  &  13.95 &  14.52  &     1     \\  
PG~1148$+$549    &  0.9754  &  0.7015  &  13.86 &  14.37  &     1     \\ 
PG~1148$+$549    &  0.9754  &  0.7248  &  13.81 &  13.86  &     1     \\ 
PKS~0405$-$123   &  0.5726  &  0.4951  &  13.96 &  14.41  &     2     \\ 
3C~263	         &   0.646  &  0.3257  &  13.98 &  13.98  &     3     \\ 
HE~0226$-$4110   &   0.495  &  0.2070  &  13.89 &  14.37  &     4     \\ 
\hline  
\hline 
\end{tabular} 
~\\ ~\\  
Note-- $^{a}$Reference (1) \citet{Meiring12}; (2) \citet{Narayanan11}; (3) 
\citet{Narayanan09,Narayanan12} (4) \citet{Savage05a}   
\label{ne8_int}  
\end{table} 
%

\subsection{Distribution of column densities} 

\begin{figure} 
\centerline{\hbox{ 
\centerline{\vbox{
\includegraphics[height=4.4cm,width=8.4cm,angle=00]{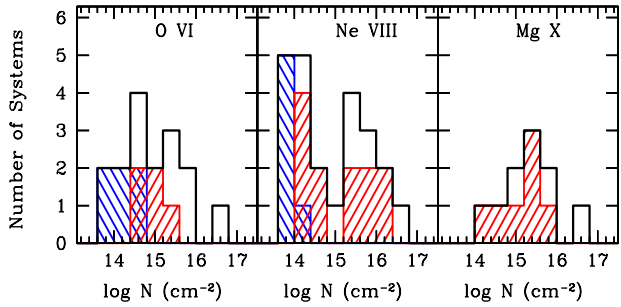} 
}}
}}
\caption{Distribution of total column densities of \os\ (left), \neo\ (middle) and 
\mgx\ (right) in all the associated \neo\ systems (i.e. using Table~\ref{tab_list} \& 
\ref{tab:summary}). The 60$^\circ$ hashed (red) histogram shows the data points from this 
paper (i.e. using Table~\ref{tab_list} only) whereas as 120$^\circ$ hashed (blue) 
histograms are for measurements in intervening systems as given in Table~\ref{ne8_int}.  
} 
\label{analysis1} 
\end{figure}

In Fig.~\ref{analysis1}, we show the column density distributions of the \os, \neo, and \mgx, as measured in intervening and associated \neo\ absorbers in our sample and from the existing literature (i.e. using Table~\ref{tab_list}, \ref{tab:summary} and \ref{ne8_int}). The (blue) 120$^\circ$ and (red) 60$^\circ$ hashed histograms show the distributions corresponding to the intervening \neo\ systems (i.e. from Table~\ref{ne8_int}) and the associated \neo\ systems from this paper (i.e. from Table~\ref{tab_list}) respectively. The histograms clearly show that the column densities of \os\ and \neo\ are systematically higher in case of associated absorbers compared to those of intervening absorbers. For example, the median values of log~$N(\neo)$ ($\rm cm^{-2}$) are 13.95$\pm$0.10 and 15.40$\pm$0.78 for the intervening and the associated absorbers respectively. The median values of log~$N(\os)$ ($\rm cm^{-2}$), on the other hand, are 14.37$\pm$0.27 and 15.40$\pm$0.58 for intervening and associated absorbers respectively. However we note that, both at high and low redshifts, \os\ absorbers do not show any compelling evidence of having different column density distribution for intervening and associated systems \citep[see e.g.][]{Tripp08,Fox08}. The apparent discrepancy is mainly because of the fact that we consider \os\ column densities measured in the \neo\ absorbers, both in the cases of intervening and associated systems. To our knowledge no \mgx\ absorption has ever been reported in intervening systems. The median value of log~$N(\mgx)$ ($\rm cm^{-2}$) in associated systems turns out to be 15.47$\pm$0.66. The median values of \os, \neo, and \mgx\ in our sample are very similar. But we caution here that we have assumed all the upper/lower limits as measurements.  

It is evident from the middle panel of Fig.~\ref{analysis1} that all the associated absorbers show $N(\neo) > 10^{14} \rm cm^{-2}$. This could primarily be due to the fact that we are not sensitive enough to detect a broad line with $N(\neo) < 10^{14} \rm cm^{-2}$ in the COS spectra used here. For example, for a typical $S/N$ ratio of $\sim 10$, the 5$\sigma$ upper limit for non-detection of \neo$\lambda$770 line is log~$N(\neo)~(\rm cm^{-2}) < 13.66$ for $b$-parameter of 100 \kms. Here the assumed $b$ value (i.e. 100 \kms) is typical for mini-BAL system. The previously reported associated \neo\ systems (see e.g. Table~\ref{tab:summary}) are all showing $N(\neo)>$ 10$^{15}$ cm$^{-2}$.  

\subsection{Column density ratios and ionization state} 
\begin{figure} 
\centerline{\hbox{ 
\centerline{\vbox{
\includegraphics[height=8.4cm,width=8.4cm,angle=00]{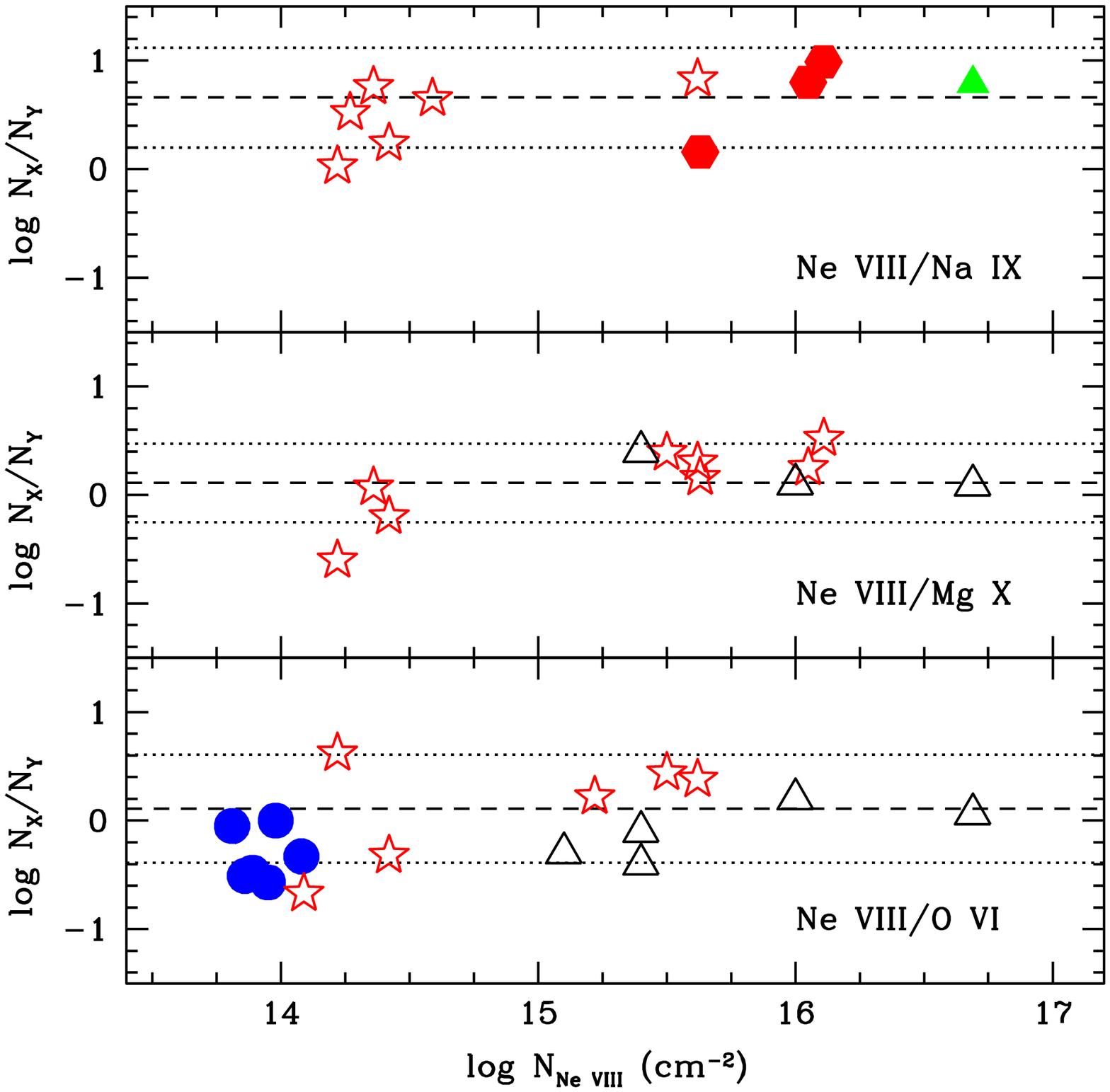} 
}}
}}
\caption{Column density ratios (\neo/\os, {\sl bottom}; \neo/\mgx, {\sl middle};  
	\neo/\nani, {\sl top}) as a function of $N(\neo)$. The open triangles 
	and (blue) filled circles are from Table~\ref{tab:summary} and Table~\ref{ne8_int}  
	respectively. In the top panel all the points apart from (green) triangle is from this 
	paper. The solid hexagon indicate \nani\ detections. The solid (green) triangle represents  
	the tentative \nani\ detection by \citet{Arav99a}. The mean values and 
	corresponding scatters, only for data points from our sample, are shown in each panel by 
	horizontal dashed and dotted lines respectively.   
	Here we assumed all the limits as measurements.      
} 
\label{analysis2} 
\end{figure}

In different sub-panels of Fig.~\ref{analysis2}, various column density ratios are plotted as a function of log~$N(\neo)$. The {\sl stars} and the {\sl circles} in the bottom panel of Fig.~\ref{analysis2} are representing the $N(\neo)/N(\os)$ ratios in associated and intervening \neo\ absorbers respectively. A Spearman rank correlation analysis shows ($\rho_s = 0.75$) a 2.3$\sigma$ level correlation between \neo\ and \os\ column densities in associated absorbers. When we include the intervening absorbers in the analysis, the correlation becomes even tighter (e.g. $\rho_s = 0.91$ and $\rho_s/\sigma$ = 3.5).  
The median value of log~$N(\neo)/N(\os)$ ratio for the associated absorbers is 0.11$\pm$0.50. Under photoionization equilibrium it corresponds to the ionization parameter log~U = 0.4$\pm$0.2. Under collisional ionization equilibrium the above ratio is reproduced when T $\sim$ $10^{5.8}$ K and log~U~$\le-$2. Based on the present data we are not in a position to disentangle among different ionization mechanisms. However, detection of absorption line variability and its relationship to the continuum variation will enable us to distinguish between the two alternatives. For a flat SED, the ionization parameter, density ($n_{\rm H}$) and distance between the absorber and the QSO are related by,  
\begin{equation} 
{\rm log} \left(\frac{n_{\rm H}}{10^{5} \rm /cc}\right) 
= {\rm log}~L_{912\rm\AA}^{30} -{\rm log} \left( \frac{r}{100 \rm pc}\right)^{2} - {\rm log~U} -1.25      
\label{eqn:nH_rC}
\end{equation} 
where, log~$L_{912\rm\AA}^{30}$ is the monochromatic luminosity of the QSO at the Lyman continuum in units of 10$^{30}$ erg s$^{-1}$ Hz$^{-1}$.  The density estimation using absorption line variability or fine-structure excitations will enable us to get the location of the absorbing gas with respect to the central engine. This will allow us to estimate the kinetic luminosity of the outflow which is very crucial for probing the AGN feedback \citep[e.g.,][]{Moe09,Dunn10,Bautista10,Borguet12b}.  

We have mentioned earlier that the intervening absorbers are showing systematically lower values of $N(\neo)$. But, as far as $N(\neo)/N(\os)$ ratio is concerned there is very little difference between associated and intervening absorbers. For example, the median values of log~$N(\neo)/N(\os)$ ratios in intervening and associated absorbers are, $-$0.45$\pm$0.50 and 0.11$\pm$0.50 respectively, consistent within 1$\sigma$ level. Naively this implies a large ionization parameter even for the intervening systems. In case of intervening \neo\ absorbers models of collisional ionization are generally proposed, as photoionization by the extragalactic UV background \citep[]{Haardt96} requires unusually large cloud sizes \citep[]{Savage05a,Narayanan09,Narayanan11}. Thus similar ratios seen between associated and intervening systems and between different components in an associated system as in the case of \citet{Muzahid12b} favour collisional ionization in the associated absorbers as well.

Further, we notice a strong correlation between $N(\neo)$ and $N(\mgx)$ (i.e. $\rho_s$ = 0.90 and $\rho_s/\sigma$ = 2.4). The median value of log~$N(\neo)/N(\mgx)$ is found to be 0.11$\pm$0.36 (see middle panel of Fig.~\ref{analysis2}). These observed ratios correspond to a very narrow range in gas temperature (i.e.~$T \sim 10^{5.95\pm0.03}$~K) under CIE or very narrow range in ionization parameter (i.e. log~U $\sim0.8\pm0.2$) under photoionization [see panel~(D) of Fig.~\ref{cloudy1}]. Given the high ionization parameter and low neutral hydrogen column density (i.e. $< 10^{14.5}$~cm$^{-2}$ in most of the cases), the predicted total hydrogen column density is too low (i.e. $N({\rm H}) < 10^{20.5}$~cm$^{-2}$) to produce significant continuum optical depth in the soft X-ray regime.       
 
In the top panel of Fig.~\ref{analysis2} we show $N(\neo)/N(\nani)$ ratio as a function of $N(\neo)$ in logarithmic scale. It is interesting to note that the three systems, where we detect \nani\ absorption (i.e. solid hexagons in the plot), are all showing log~$N(\neo) \gtrsim$ 15.60. The solid (green) triangles in this panel represents the tentative detection of \nani\ reported by \citet{Arav99a}, which also shows log~$N(\neo)>$15.60. We notice that log~$N(\neo)/N(\mgx)$  and  log~$N(\neo)/N(\os)$ are roughly similar between the systems with and without detectable \nani\ absorption. This clearly means that the lack of \nani\ detection can be attributed to low $N\rm (H)$.  

%
%

\subsection{Multiple phases in \nani\ absorbers} 

We report secure detections of \nani\ absorption in three associated \neo\ systems for the first time. These systems show signatures of multiple component structure. Photoionization models with log~U $\sim$ 1 explain the \nani\ phase of the absorbers. However these models require $\rm Na$ abundance being enhanced by a factor of 4--7 with respect to $\rm Mg$. Standard chemical evolution models do not predict such large enhancement of $\rm Na$ over $\rm Mg$ \citep[see Fig. 18 of][and Fig. 6 of \citet{Venn04}]{Timmes95}. The photoionization models also suggest a typical density of the absorbing region varying by up to a factor 10 along the transverse direction. As photoionization predicts roughly same temperature for the range of ionization parameters probed by low and high ions, different phases cannot be in pressure equilibrium. On the contrary, if collisional excitations are important then one may not need an enhancement of $\rm Na$, provided the gas temperature $T = 10^{5.9}$ K. In the case of CIE the low ionization phase requires $T\sim 10^{5.2}$ K. Therefore, a factor $\sim$~5 density difference between two phases is needed for the gas to be in pressure equilibrium. In the case of CIE, absorbing gas has to be far away from the QSO for the gas to be unaffected by the QSO radiation. Therefore it is important to identify the source of energy that maintains the high temperature of the gas. Probing the optical depth variability and presence of fine-structure transitions with new $HST/$COS observations will allow us to make good progress in this direction. 

At last, we note that the element $\rm Na$ has not been incorporated in the non-equilibrium collisional ionization calculations so far. For the metallicity as measured in our sample, non-equilibrium effects would be important and  can provide more realistic models of \nani\ absorbers. Therefore, inclusion of $\rm Na$ in non-equilibrium calculations will be very useful.

\section{Summary \& Conclusions}  
\label{con} 

We present a sample of new class of associated absorbers, detected through \neo$\lambda\lambda$770,780 absorption, in $HST/$COS spectra of intermediate redshift (0.45~$\le z \le$~1.21) quasars. We searched for \neo\ absorption in the public $HST/$COS archive of QSOs with $S/N \ge 10$ and emission redshift \zem~$ > 0.45$. There were total 20 QSO sight lines in the $HST/$COS archive before February 2012, satisfying these criteria. Seven of these QSOs are radio bright. The signatures of associated \neo\ absorption are seen in 40\% (i.e. 8 out of 20) of the lines of sight, with 10 secured and 2 tentative \neo\ systems detected in total. None of them are towards radio bright QSOs. The associated absorbers detected towards QSO HE~0226$-$4110 and QSO HE~0238$-$1904 were previously reported by \citet{Ganguly06} and \citet{Muzahid12b} respectively. Here we summarize our main results.  

\vskip 0.2cm \noindent {\bf (1)}   
Majority of the \neo\ absorbers are detected with outflow velocities $\lesssim$5000 \kms. The highest velocity system shows $|v_{\rm ej}| \sim 19,000$~\kms. Medium resolution COS spectra allow us to probe the component structure of \neo\ absorption in most of the systems. The line spread of \neo\ absorption is found to be in the range 100~$\le \delta v (\rm km~s^{-1}) \le$~1000, suggesting that these absorbers are most likely  mini-BALs. The Doppler parameters measured in individual components (with median 58.7$\pm$31.7 \kms) indicates domination of non-thermal motions.             

\vskip 0.2cm \noindent {\bf (2)}   
We detect \mgx\ absorption in 7 of 8 \neo\ systems when the lines are not blended and are covered by the observations. Moreover, we report first secure detections of \nani\ absorption in three highest velocity systems in our sample. All three \nani\ systems show high $N(\neo)$ (i.e.$ > 10^{15.6}$ cm$^{-2}$). The measurements and/or limits on the column densities of different ions, detected in these \nani\ absorbers, require very high ionization parameter (i.e. log~U $\ge 0.5$) and high metallicity (i.e. $Z \ge Z_{\odot}$) when we consider single phase photoionization models. However, ionization potential dependent covering fraction seen in these absorbers suggests kinematic coincidence of multiphase gas with higher ionization species having higher projected area. Given the high value of ionization parameter (log~U) and observed low $N(\hi)$, the model predicted $N(\rm H)$ is too low (i.e. $<10^{20.5}$~cm$^{-2}$) to produce any significant continuum optical depth in the soft X-ray regime. The observed $N(\mgx)/N(\nani)$ ratios, under single phase photoionization scenario, require a factor $\gtrsim 5$ enhancement of $\rm Na$ abundance with respect to $\rm Mg$. However, such enhancement is not required in CIE models provided gas temperature is $T \ge 10^{5.9}$~K. In the case of CIE, the low ions require a different phase with temperature $T \sim 10^{5.2}$ K suggesting a factor of $\sim 5$ difference in density between two gas phases to be in pressure equilibrium.             

\vskip 0.2cm \noindent {\bf (3)} 
We notice a very narrow range in the column density ratios of high ions (i.e. \os, \neo, \mgx\ etc.). This suggests a narrow range in ionization parameter (temperature) under photoionization (CIE). The median value of log~$N(\neo)/N(\os)$ $= 0.11\pm0.50$ as measured in our sample is comparable to that measured in the intervening \neo\ absorbers within the measurement uncertainties. In case of intervening \neo\ absorbers collisional ionization is generally proposed, as photoionization by the extragalactic UV background requires unusually large cloud sizes. Indeed, CIE can play an important role in deciding the ionization structure of the absorbing gas in our sample as well. However, for CIE to be dominant, gas cloud has to be far away from the QSO. In that case it is crucial to understand sources of thermal and mechanical energy and the stability of the absorber. Variability study with repeated $HST/$COS observation is needed to make further progress on these issues.     

\section{acknowledgment} 

We thank anonymous referee for useful comments. We appreciate the efforts of the people involved with the design and construction of COS and its deployment on the $HST$. Thanks are also extended to the people responsible for determining the orbital performance of COS and developing the {\sc calcos} data processing pipeline. We thankfully acknowledge Dr. Jane Charlton for providing the STIS E230M spectrum of PG~1206$+$459. We thank Dr. Gulab C. Dewangan and Dr. Durgesh Tripathi for useful discussions. SM thanks Sibasish Laha for useful discussions on {\sc cloudy} modelling. SM also thanks CSIR for providing support for this work. RS wish to thank Indo-French Centre for the Promotion of Advanced Research under the programme No. 4304--2. NA acknowledge support from NASA STScI grants AR-12653.  

\def\aj{AJ}%
\def\actaa{Acta Astron.}%
\def\araa{ARA\&A}%
\def\apj{ApJ}%
\def\apjl{ApJ}%
\def\apjs{ApJS}%
\def\ao{Appl.~Opt.}%
\def\apss{Ap\&SS}%
\def\aap{A\&A}%
\def\aapr{A\&A~Rev.}%
\def\aaps{A\&AS}%
\def\azh{AZh}%
\def\baas{BAAS}%
\def\bac{Bull. astr. Inst. Czechosl.}%
\def\caa{Chinese Astron. Astrophys.}%
\def\cjaa{Chinese J. Astron. Astrophys.}%
\def\icarus{Icarus}%
\def\jcap{J. Cosmology Astropart. Phys.}%
\def\jrasc{JRASC}%
\def\mnras{MNRAS}%
\def\memras{MmRAS}%
\def\na{New A}%
\def\nar{New A Rev.}%
\def\pasa{PASA}%
\def\pra{Phys.~Rev.~A}%
\def\prb{Phys.~Rev.~B}%
\def\prc{Phys.~Rev.~C}%
\def\prd{Phys.~Rev.~D}%
\def\pre{Phys.~Rev.~E}%
\def\prl{Phys.~Rev.~Lett.}%
\def\pasp{PASP}%
\def\pasj{PASJ}%
\def\qjras{QJRAS}%
\def\rmxaa{Rev. Mexicana Astron. Astrofis.}%
\def\skytel{S\&T}%
\def\solphys{Sol.~Phys.}%
\def\sovast{Soviet~Ast.}%
\def\ssr{Space~Sci.~Rev.}%
\def\zap{ZAp}%
\def\nat{Nature}%
\def\iaucirc{IAU~Circ.}%
\def\aplett{Astrophys.~Lett.}%
\def\apspr{Astrophys.~Space~Phys.~Res.}%
\def\bain{Bull.~Astron.~Inst.~Netherlands}%
\def\fcp{Fund.~Cosmic~Phys.}%
\def\gca{Geochim.~Cosmochim.~Acta}%
\def\grl{Geophys.~Res.~Lett.}%
\def\jcp{J.~Chem.~Phys.}%
\def\jgr{J.~Geophys.~Res.}%
\def\jqsrt{J.~Quant.~Spec.~Radiat.~Transf.}%
\def\memsai{Mem.~Soc.~Astron.~Italiana}%
\def\nphysa{Nucl.~Phys.~A}%
\def\physrep{Phys.~Rep.}%
\def\physscr{Phys.~Scr}%
\def\planss{Planet.~Space~Sci.}%
\def\procspie{Proc.~SPIE}%
\let\astap=\aap
\let\apjlett=\apjl
\let\apjsupp=\apjs
\let\applopt=\ao
\bibliographystyle{mn}
\bibliography{mybib}

\end{document}